\documentclass[12pt]{article}
\pdfoutput=1
\usepackage{amsmath}
\usepackage{amssymb}
\usepackage{epsfig}
\usepackage{cite}
\usepackage{fontenc}
\usepackage{graphicx}
\usepackage{float}
\usepackage[lofdepth,lotdepth,caption=false]{subfig}
\usepackage{color}
\usepackage{booktabs}
\usepackage{tabularx}
\usepackage{multirow}
\usepackage{longtable}
\usepackage{lscape}
\usepackage{bbm}
\usepackage{dcolumn}
\usepackage{rotating}

\usepackage[normalem]{ulem}
\definecolor{darkgreen}{cmyk}{1,0,1,0.4}
\definecolor{brown}{cmyk}{0,0.8,1,0.2}
\definecolor{darkred}{cmyk}{0,1,1,0.2}

\allowdisplaybreaks[4] 
\makeatletter
\renewcommand{\fnum@table}{\textbf{\tablename~\thetable}}
\renewcommand{\fnum@figure}{\textbf{\figurename~\thefigure}}
\makeatother

\newcounter{myenumi}

\renewcommand{\themyenumi}{\roman{myenumi}}
{\end{list}}

\newlength{\myem}
\newcounter{mysubequation}[equation]

\makeatletter
\renewcommand{\section}{\@startsection{section}{1}{0em}{-\baselineskip}%
{\baselineskip}{\normalfont\large\bfseries}}
\renewcommand{\subsection}%
{\@startsection{subsection}{2}{0em}{-0.7\baselineskip}%
{0.7\baselineskip}{\normalfont\bfseries}}
\makeatother

\newcommand{\bi}{\begin{itemize}}
\newcommand{\ei}{\end{itemize}}

\def\beq{\begin{equation}}
\def\eeq{\end{equation}}

\newcommand{\bea}{\begin{eqnarray}}
\newcommand{\eea}{\end{eqnarray}}

\newcommand{\ie}{{\it i.e.}}

\newcommand{\eg}{{\it e.g.}}

\newcommand{\etc}{{\it etc.}}

\newcommand{\eet}{\varepsilon_{e\tau}}
\newcommand{\emt}{\varepsilon_{\mu\tau}}
\newcommand{\ett}{\varepsilon_{\tau\tau}}

\newcommand{\eem}{\varepsilon_{e\mu}}
\newcommand{\eme}{\varepsilon_{\mu e}}
\newcommand{\emm}{\varepsilon_{\mu\mu}}

\newcommand{\nue}{\ensuremath{\nu_e}}
\newcommand{\numu}{\ensuremath{\nu_\mu}}

\def\vareps{\varepsilon}
\newcommand{\chisq}{\ensuremath{\chi^2}}

\newcommand{\liar}{LAr}

\newcommand\schd{Schr$\ddot{\rm o}$dinger}
\newcommand\sch{Schr$\ddot{\rm o}$dinger~}

\def\<{\langle}
\def\>{\rangle}

\def\dfrac#1#2{{\displaystyle\frac{#1}{#2}}}

\def\lsim{\mathrel{\rlap{\lower4pt\hbox{\hskip1pt$\sim$}}
    \raise1pt\hbox{$<$}}}         
\def\gsim{\mathrel{\rlap{\lower4pt\hbox{\hskip1pt$\sim$}}
    \raise1pt\hbox{$>$}}}         

\begin{document}
%

\begin{titlepage}

\renewcommand{\thefootnote}{\alph{footnote}}

\vspace*{-3.cm}
\begin{flushright}

\end{flushright}


\renewcommand{\thefootnote}{\fnsymbol{footnote}}
\setcounter{footnote}{-1}

{\begin{center}
{\large\bf
 Testing non-standard neutrino matter interactions in 
 atmospheric neutrino propagation \\[0.2cm]
}
\end{center}}

\renewcommand{\thefootnote}{\alph{footnote}}

\vspace*{.8cm}
\vspace*{.3cm}
{\begin{center} 
{{\sf  
				Animesh Chatterjee$^\star$~\footnote[1]{\makebox[1.cm]{Email:}
                animesh.chatterjee@uta.edu},
                Poonam Mehta$^{\dagger}$~\footnote[2]{\makebox[1.cm]{Email:}
                pm@jnu.ac.in}, 
                Debajyoti Choudhury$^\diamond$~\footnote[3]{\makebox[1.cm]{Email:}
                debchou@physics.du.ac.in} and 
                Raj Gandhi$^\ddagger $~\footnote[4]{\makebox[1.cm]{Email:}
                raj@hri.res.in}

               }}

\end{center}}
\vspace*{0cm}
{\it 
\begin{center}

$^\star$\, Department of Physics, University of Texas at Arlington,
Arlington, TX 76019, USA 

 $^\dagger$ \,     School of Physical Sciences, 
        Jawaharlal Nehru University, 
      New Delhi 110067, India

 $^\diamond$ \,   Department of Physics and Astrophysics, 
        University of Delhi, 
      Delhi 110007, India

$^\ddagger$\,	Harish-Chandra Research Institute, Chhatnag Road, Jhunsi, Allahabad 211 019, India



\end{center}}

\vspace*{1.5cm}

\begin{center}
{\Large \today}
\end{center}

{\Large 
\bf
\begin{center} Abstract 
\end{center} 
 }

 We study the effects of non-standard interactions on the oscillation
 pattern of atmospheric neutrinos. We use neutrino oscillograms as our
 main tool to infer the role of non-standard interactions (NSI)
 parameters at the probability level in the energy range, $E \in
 [1,20]$ GeV and zenith angle range, $\cos \theta \in [-1,0]$. We
 compute the event rates for atmospheric neutrino events in presence
 of NSI parameters in the energy range $E \in [1,10]$ GeV for two
 different detector configurations - a magnetized iron calorimeter and
 an unmagnetized liquid Argon time projection chamber which have
 different sensitivities to NSI parameters due to their complementary
 characteristics. 

\vspace*{.5cm}

\end{titlepage}

\newpage

\renewcommand{\thefootnote}{\arabic{footnote}}
\setcounter{footnote}{0}

\section{Introduction}

 With the immense progress over the past few decades in establishing
 neutrino masses and mixings, it is fair to say that neutrino physics
 has entered an era of precision measurements.  The first confirmation
 came in 1998 courtesy the pioneering experiment, Super Kamiokande
 (SK)~\cite{Fukuda:1998mi}. With more data as well as with the aid of
 numerous other experiments, we have steadily garnered more and more
 precise information about the neutrino mixing parameters. As a
 result, the long list of unanswered questions in the standard
 scenario has become shorter (for recent global analyses of all
 neutrino oscillation data, see
 \cite{GonzalezGarcia:2012sz,Capozzi:2013csa,Forero:2014bxa}). The
 focus of the ongoing and future neutrino experiments is on resolving
 the issue of neutrino mass hierarchy \ie, sign $(\delta
 m^2_{31})$\footnote{$\delta m^2 _{31} = m^2 _{3} - m^2_1$. },
 measuring the CP phase ($\delta$) and determining the correct octant
 of the mixing angle $\theta_{23}$.

The minimal theoretical scenario needed to describe oscillations
requires the existence of neutrino masses.  The simplest way is to add
right handed neutrino fields to the Standard Model (SM) particle
content (something that the originators of the SM would, no doubt,
have trivially done were nonzero neutino masses known then) and
generate a Dirac mass term for neutrinos. However it is hard to
explain the smallness of the neutrino mass terms via this
mechanism. To overcome this, an attractive way is to add
dimension-five non-renormalizable terms consistent with the symmetries
and particle content of the SM, which naturally leads to desired tiny
Majorana masses for the left-handed neutrinos\footnote{{Terms
such as $\overline{\nu_{R i}^c} \nu_{R j}^{}$ are gauge invariant too and
phenomenologically unconstrained. While they break lepton number, the
latter is only an accidental symmetry in the SM. Thus, such terms, in
conjunction with the usual Dirac mass terms, would generate tiny
observable neutrino masses through the see-saw mechanism. It can be
readily seen that, the aforementioned dimension-five term
($\overline{L^c_{}} L \phi \phi$) essentially mimics this mechanism in
the low energy effective theory.}}. However in the minimal scenario of
this extension, the dominant neutrino interactions involving the light
fields are still assumed to be described by weak interactions within
the SM in which flavour changes are strongly suppressed.

Once we invoke new physics in order to explain the non-zero neutrino
masses, it seems rather unnatural to exclude the possibility of non
standard interactions (NSI) which can, in principle, allow for flavour
changing interactions. Simultaneously, these are new sources of CP
violation which can affect production, detection and propagation of
neutrinos~\cite{Ohlsson:2012kf}. Some of the early attempts discussing
new sources of lepton flavour violation (for instance, $R$-parity
violating supersymmetry) were geared towards providing an alternate
explanation for the observed deficit of neutrinos in the limiting case
of zero neutrino masses and the absence of vacuum
mixing~\cite{Grossman:1995wx,PhysRevLett.82.3202}. In recent years,
the emphasis has shifted towards understanding the interplay between
the standard electroweak interactions (SI) and NSI
and whether future oscillation experiments can test such NSI apart
from determining the standard oscillation parameters precisely. This
has led to an upsurge in research activity in this area (see the
references in~\cite{Ohlsson:2012kf}).
This is also interesting from the point of view of
complementarity with the collider searches for new physics. There are
 other motivations for NSI as well such as
(electroweak) leptogenesis~\cite{Pilaftsis:2005rv}, neutrino magnetic
moments~\cite{Barbieri:1988fh,Babu:1989wn,Choudhury:1989pw,Healey:2013vka},
neutrino condensate as dark energy~\cite{Bhatt:2009wb,Wetterich:2014gaa}. 
 
Neutrino oscillation experiments can probe NSI by exploiting the interference with the Standard Model amplitude.  In view of the
excellent agreement of data with standard flavour conversion via
oscillations, we would like to explore the extent to which NSI
(incorporated into the Lagrangian phenomenologically via small
parameters) is empirically viable, with specific focus on atmospheric
neutrino signals in future detectors. NSI in the context of
atmospheric neutrinos has been studied by various
authors~\cite{Fornengo:2001pm,Huber:2001zw,GonzalezGarcia:2004wg,GonzalezGarcia:2011my,Esmaili:2013fva}. Also there are studies pertaining to other new physics scenarios using atmospheric neutrinos such as CPT violation~\cite{Datta2004356,Chatterjee:2014oda}, violation of the equivalence principle~\cite{Esmaili:2014ota}, large extra dimension models~\cite{Esmaili:2014esa} and sterile neutrinos~\cite{Esmaili:2012nz,Esmaili:2013cja,Esmaili:2013vza}.

The plan of the article is as follows. We first briefly outline the
NSI framework in Sec.~\ref{sec:framework} and 
{subsequently} discuss the neutrino oscillation probabilities in
presence of NSI using the perturbation theory approach (in
Sec.~\ref{sec:prob}). We describe the features of the neutrino
oscillograms in Sec.~\ref{sec:oscillograms}. 
We give the details of our analysis in 
Sec.~\ref{sec:analysis} and the discussion
on events generated for the two detector types in
Sec.~\ref{sec:events}. Finally, we conclude in
Sec.~\ref{sec:conclude}.

\section{Neutrino NSI Framework: relevant parameters and present constraints} 
\label{sec:framework}

As in the case of standard weak interactions, a wide class of ``new
physics scenarios" can be conveniently parameterised in a model
independent way at low energies ($E \ll M_{EW}$, where $M_{EW}$ is the
electroweak scale) by using effective four-fermion interactions.  In
general, NSI can impact the neutrino oscillation signals via two kinds
of interactions : (a) charged current (CC) interactions (b) neutral
current (NC) interactions.  However, CC interactions affect processes
only at the source or the detector and these are are clearly
discernible at near detectors (see for example,
\cite{Datta:2000ci,Mehta:2001na}).  On the other hand, the NC
interactions affect the propagation of neutrinos which can be studied
 {only} at far detectors. Due to this decoupling, the two can be
treated in isolation. Usually, it is assumed that the CC NSI terms
(\eg, of the type $(\bar \nu_\beta \gamma^\mu P_L l_\alpha)(\bar f_{L}
\gamma_\mu P_C f'_{L})$ with $f,f'$ being the components of a weak
doublet) are more tightly constrained than the NC terms and, hence,
are not considered.  It turns out, though, that, in specific models,
the two can be of comparable strengths~\cite{Biggio:2009nt}.  However,
since we are interested in NSI that alter the propagation of
neutrinos, we shall consider the NC type of interactions alone.

The effective Lagrangian describing the NC type neutrino NSI of the
type $(V-A)(V \pm A)$ is given by\footnote{One could think that other Dirac
  structures generated by intermediate scalar ($S$), pseudoscalar
  ($P$) or tensor ($T$) fields may also be there. However, these
  would only give rise to subdominant effects.}
\begin{equation}
\label{nsilag}
{\cal L}_{NSI} = -2 \sqrt 2 G_F \varepsilon_{\alpha \beta}^{f\, C} ~ [\bar \nu_\alpha \gamma^\mu P_L \nu_\beta] ~[\bar f \gamma_\mu P_C f]~,
\end{equation}
where $G_F$ is the Fermi constant, $\nu_{\alpha},\nu_{\beta}$ are
neutrinos of different flavours, and $f$ is a first generation SM
fermion ($e,u,d$)~\footnote{Coherence requires that the flavour of the
background fermion ($f$) is preserved in the interaction.  Second or
third generation fermions do not affect oscillation experiments since
matter does not contain them.}.  The chiral projection operators are
given by $P_L = (1 - \gamma_5)/2$ and $P_C=(1\pm \gamma
_5)/2$. {If} the NSI {arises} at
scale $M_{NP} \gg M_{EW}$ from some higher dimensional operators (of
order six or higher),  {it
would imply} a suppression of at least $\varepsilon_{\alpha\beta}^{fC}
\simeq (M_{EW}/M_{NP})^2$ (for $M_{NP} \sim 1~TeV$, we have
$\varepsilon_{\alpha\beta}^{fC} \simeq 10^{-2}$).  {However, such a
naive dimensional analysis argument breaks down if the new physics
sector is strongly interacting as can happen in a variety of
models. We shall, hence, admit even larger
$\varepsilon_{\alpha\beta}^{fC}$ as long as these are consistent with all
current observations.}  In general, NSI terms can be complex.
Naively, $SU(2)$ invariance would dictate that operators involving
$\nu_{Li}$ must be accompanied by ones containing the corresponding
charged lepton field, thereby leading to additional CC
interactions. This, however, can be avoided by applying to $SU(2)$
breaking and/or invoking multiple fields and interactions in the heavy
(or hidden) sector. Rather than speculate about the origin of any such
mechanism, we assume here (as in much of the literature) that no such
CC terms exist.

  The new NC interaction terms can affect the
neutrino oscillation physics either by causing the flavour of neutrino
to change ($\nu_\alpha + f \to \nu_\beta + f$) \ie, flavour changing
(FC) interaction or, by having a non-universal scattering amplitude of
NC for different neutrino flavours \ie, flavour preserving (FP)
interaction.  At the level of the underlying Lagrangian, NSI coupling
of the neutrino can be to $e,u,d$. However, from a phenomenological
point of view, only the sum (incoherent) of all these individual
contributions (from different scatterers) contributes to the coherent
forward scattering of neutrinos on matter. If we normalize\footnote{If
  we normalize to either up or down quark abundance (assume isoscalar
  composition of matter) instead, there is a relative factor of 3
  which will need to be incorporated accordingly.} to $n_e$, the
effective NSI parameter for neutral Earth matter\footnote{For neutral
  Earth matter, there are 2 nucleons (one proton and one neutron) per
  electron. For neutral solar matter, there is one proton for one
  electron, and $\varepsilon_{\alpha\beta} = \varepsilon ^e_{\alpha\beta} +
  2\varepsilon^u_{\alpha \beta} + 2 \varepsilon^d_{\alpha\beta}$} is 
\bea
\varepsilon_{\alpha\beta} &=& \sum_{f=e,u,d} \dfrac{n_f}{n_e}
\varepsilon_{\alpha\beta}^f = \varepsilon_{\alpha\beta}^e +2
\varepsilon_{\alpha\beta}^u + \varepsilon_{\alpha\beta}^d + \dfrac{n_n}{n_e}
(2\varepsilon_{\alpha\beta}^d + \varepsilon_{\alpha\beta}^u) = \varepsilon
^e_{\alpha\beta} + 3 \varepsilon^u_{\alpha \beta} + 3
\varepsilon^d_{\alpha\beta} \ ,
      \label{eps_combin}
\eea 
where $n_f$ is the density of fermion $f$ in medium crossed by the
neutrino and $n$ refers to neutrons.  Also, $\varepsilon_{\alpha\beta}^f=
\varepsilon_{\alpha\beta}^{fL} + \varepsilon_{\alpha\beta}^{fR}$ which
encodes the fact that NC type NSI matter effects are sensitive to the
vector sum of NSI couplings.

Let us, now, discuss the constraints on {the} NC type NSI parameters.  As
mentioned above, the combination that
enters oscillation physics is given by
Eq.~(\ref{eps_combin}). The individual NSI terms such as
$\varepsilon_{\alpha \beta}^{f L}$ or $\varepsilon_{\alpha \beta}^{f R}$ are
constrained in any experiment (keeping only one of them non-zero at a
time) and moreover the coupling is either to $e,u,d$
individually~\cite{Davidson:2003ha}. In view of this, it is not so
straightforward to interpret those bounds in terms of {an} effective
$\varepsilon_{\alpha \beta}$. There are two ways : (a) One could take a conservative approach and use the
most stringent constraint in the individual NSI terms (say, use 
$|\varepsilon_{\mu \tau}^{u}|$) to constrain the effective term
(say, $|\varepsilon_{\mu\tau}|$) in Eq.~(2) and that leads to
 \begin{eqnarray}
 |\vareps_{\alpha\beta}|
 \;<\;
  \left( \begin{array}{ccc}
0.06  &
0.05 & 
0.27 \\
0.05 &0.003 & 0.05 \\
0.27  &
0.05 &
0.16 \\
  \end{array} \right) \ . \label{smallnsi}
\end{eqnarray} 
The constraints involving muon neutrinos are at least an order of
magnitude stronger (courtesy the NuTeV and CHARM
scattering experiments) than those involving electron and tau
neutrino~\cite{Escrihuela:2011cf}.  (b) With the assumption that the
errors on individual NSI terms are uncorrelated, the authors in
Ref.~\cite{Biggio:2009nt} 
deduce model-independent bounds on effective NC NSI terms
\begin{eqnarray}
\varepsilon_{\alpha\beta} \lsim \left\{  \sum_{C=L,R} [ (\varepsilon_{\alpha \beta}^{e C} )^2 + (3 \varepsilon_{\alpha\beta}^{u C})^2 + (3 \varepsilon _{\alpha \beta}^{d C})^2 ] \right\}^{1/2} \ ,
\end{eqnarray} 
 which,  for neutral Earth matter,  leads to
 \begin{eqnarray}
 |\varepsilon_{\alpha\beta}|
 \;<\;
  \left( \begin{array}{ccc}
4.2  &
0.33 & 
3.0 \\
0.33 &0.068 & 0.33 \\
3.0  &
0.33 &
21 \\
  \end{array} \right) \ .\label{largensi}
\end{eqnarray} 
 Note that the values mentioned in Eq.~(\ref{largensi}) are
larger by one or two orders of magnitude than the {overly restrictive} bounds of Eq.~(\ref{smallnsi}), which,
of course, need not be applicable.
 
Apart from the model independent theoretical bounds, two experiments
have used the neutrino data to constrain NSI parameters which are more restrictive. The SK NSI search in atmospheric neutrinos crossing the Earth found no evidence
in favour of NSI and the study led to upper bounds on NSI
parameters~\cite{Mitsuka:2011ty} given by $|\varepsilon_{\mu\tau}| < 0.033, |
\varepsilon_{\tau\tau} - \varepsilon_{\mu\mu} | < 0.147 $ (at 90\% CL) in a
two flavour hybrid model~\cite{Ohlsson:2012kf}\footnote{The SK
collaboration uses a different normalization ($n_d$) while writing the
effective NSI parameter (see Eq.~(\ref{eps_combin})) and hence we need
to multiply the bounds mentioned in Ref.~\cite{Mitsuka:2011ty} by a
factor of 3.}.  The off-diagonal NSI parameter
$\varepsilon_{\mu\tau}$ is constrained $-0.20 < \varepsilon_{\mu\tau} <
0.07$ (at 90\% CL) from MINOS data in the framework of two flavour
neutrino oscillations~\cite{Adamson:2013ovz,Kopp:2010qt}. 
%
%
 %
 It should be noted, though, that the derivation of these bounds (the
SK one in particular~\cite{Mitsuka:2011ty}) hinge upon certain assumptions. The primary
theoretical assumption relates to the simplification of the system
onto a (hybrid) two-flavour scenario. Within the SM paradigm, this
approximation is expected to be a very good one. The situation changes
considerably, though, once NSI are introduced. As we shall see
shortly, the major effect of NSI accrues through matter effects (even
in the limit of the $\nu_e$ decoupling entirely). However, there
exists a nontrivial interplay between such effects and the
corresponding matter effects induced by canonical three-flavour
oscillations. Consequently, approximations pertaining to the
neutrino mixing matrix can significantly alter conclusions reached
about NSI. Similarly, the very presence of NSI can leave its imprint 
in the determination of neutrino parameters. A second set of imponderables relate to statistical and systematic
uncertainties, including but not limited to earth density and
atmospheric neutrino profiles. 
Thus, it is quite conceivable that the
constraints quoted by the SK collaboration could be relaxed to a fair
degree, though perhaps not to the extent of those in Eq.~(\ref{largensi}). In view
of this, and following several other studies~\cite{Ohlsson:2013epa}, we will use a
value of $|\varepsilon_{\alpha \beta}| = 0.15$ (for the parameters
$\varepsilon_{\mu\tau}$, $\varepsilon_{e\mu}$ and $\varepsilon_{e\tau}$) in our
oscillogram diagrams.  This value is eminently in agreement with Eq.~(\ref{largensi}).  Note, though, that this choice is essentially
to aid visual appreciation of the differences in the oscillogram
structures wrought by NSI. Indeed, the experimental sensitivities 
that we shall be deriving are comparable to (and often significantly 
better than) those achieved by the SK collaboration. Furthermore, we shall 
not be taking recourse to two-flavour simplifications to reach 
such sensitivities. 
Additionally, the allowed ranges of NSI parameters have been recently extracted using global analysis of neutrino data in Ref.~\cite{Gonzalez-Garcia:2013usa}. 

\section{Neutrino oscillation probability in matter with NSI}
\label{sec:prob}

The purpose of the analytic expressions presented here is to
understand the features in the probability in the presence of
NSI. All the plots presented in this paper are obtained numerically by
solving the full three flavour neutrino propagation equations
 using the PREM density profile of the Earth, and
the latest values of the neutrino parameters as obtained from global
fits (see Table~\ref{tab_osc_param_input}).
\begin{table}[htb!]
 \begin{center}
  \begin{tabular}{lccc}
     \hline
 
  & & & \\
\quad Oscillation Parameter \quad & \quad Best-fit value \quad & \quad $3\sigma$ range \quad & \quad Precision (\%) \quad \\
  & & & \\
  
  \hline
  $\sin^2 \theta_{12}/10^{-1}$ & 3.23 & 2.78 - 3.75 & 14.85 \\
 
  $\sin^2 \theta_{23}/10^{-1}$ (NH) & 5.67 (4.67)$^a$ & 3.92 - 6.43 & 24.25 \\
   $\sin^2 \theta_{23}/10^{-1}$ (IH) & 5.73 & 4.03 - 6.40 & 22.72 \\
  $\sin^2 \theta_{13}/10^{-2}$ (NH) & 2.34 & 1.77 - 2.94 & 24.84 \\
	$\sin^2 \theta_{13}/10^{-2}$ (IH) & 2.40 & 1.83 - 2.97 & 23.75  \\		 
   
  $\delta m_{21}^2 \: [10^{-5}~\rm eV^2]$  & 7.60  & 7.11 - 8.18 & 7.00 \\
  
  $|\delta m_{31}^2| \: [10^{-3}~\rm eV^2] $ (NH) & 2.48  & 2.30 - 2.65 & 7.07 \\
		$|\delta m_{31}^2| \: [10^{-3}~\rm eV^2] $ (IH) & 2.38  & 2.30 - 2.54 & 5.00 \\
		$\delta /\pi $ (NH) & 1.34  & 0.0 - 2.0 & - \\
			 $\delta /\pi $ (IH) & 1.48  & 0.0 - 2.0  & - \\
  
  \hline
  \end{tabular}
  \end{center}
  \begin{quote}
  $^a$This is a local minimum in the first octant of $\theta_{23}$
  with $\Delta \chi^2 = 0.28$ with respect to {the} global minimum.
  \end{quote}
\caption{Best-fit {values} and {the} $3 \sigma$ ranges for the
  oscillation parameters used in our
  analysis~\cite{Forero:2014bxa}. Also given is the precision which is
  defined as ratio (in percentage) of the difference of extreme values
  to the sum of extreme values of parameters in the $3\sigma$
  range. Here NH (IH) refer to normal (inverted) hierarchy.}
  \label{tab_osc_param_input}

\end{table} 

The analytic computation of probability expressions in presence of
SI~\cite{Barger:1980tf,Cervera:2000kp,Gandhi:2004md,Indumathi:2004kd,Choudhury:2004sv,Gandhi:2004bj,Akhmedov:2004ny}
as well as
NSI~\cite{Ribeiro:2007ud,Kopp:2007ne,Blennow:2008eb,Kikuchi:2008vq,Meloni:2009ia,Asano:2011nj,Coloma:2011rq,Ohlsson:2013epa}
has been carried out for different experimental settings by various
authors. Note that, for atmospheric neutrinos, one can safely neglect
the smaller mass squared difference $\delta m^2_{21}$ in comparison to
$\delta m^2_{31}$ since $\delta m^2_{21} L/4 E \ll 1$ for a large
range of values of $L$ and $E$ (especially above a GeV).  This ``one
mass scale dominant'' (OMSD) approximation allows for a relatively
simple exact analytic  {formula} for the
probability (as a function of only three parameters
$\theta_{23},\theta_{13}$ and $\delta m^2_{31}$) for the case of
constant density matter~\cite{Gandhi:2004bj} with no approximation on
$s_{13}$, and it works quite well\footnote{This approximation breaks down if the value of
  $\theta_{13}$ is small since the terms containing $\delta m^2_{21}$
  can be dropped only if they are small compared to the leading order
  term which contain $\theta_{13}$. After the precise measurement of
  the value of $\theta_{13}$ by reactor experiments, this
  approximation is well justified. For multi-GeV neutrinos, this
  condition ($L/E \ll 10^{4} $ km$/$GeV) is violated for only a small
  fraction of events with $E \simeq 1$ GeV and $L \ge 10^{4}$ km.}.
In order to systematically take into account the effect of small
parameters, the perturbation theory approach is used. We review the
necessary formulation for calculation of probabilities that affect the
atmospheric neutrino propagation using the perturbation theory
approach~\cite{Ohlsson:2013epa}.
 
In the ultra-relativistic limit, the neutrino propagation is governed
by a \schd-type equation (see~\cite{raffeltbook}) with an effective
Hamiltonian
\bea 
{\mathcal
H}^{}_{\mathrm{}} &=& {\mathcal
H}^{}_{\mathrm{vac}} + {\mathcal
H}^{}_{\mathrm{SI}} + {\mathcal
H}^{}_{\mathrm{NSI}} \ ,
\eea 
where ${\mathcal H}^{}_{\mathrm{vac}} $ is the vacuum Hamiltonian and
${\mathcal H}^{}_{\mathrm{SI}}, {\mathcal H}^{}_{\mathrm{NSI}}$ are
the effective Hamiltonians in presence of 
{SI alone and NSI} respectively. {Thus,}
 \bea
 \label{hexpand} 
 {\mathcal
H}^{}_{\mathrm{}} &=& 
\dfrac{1}{2 E} \left\{ {\mathcal U} \left(
\begin{array}{ccc}
0   &  &  \\  &  \delta m^2_{21} &   \\ 
 &  & \delta m^2_{31} \\
\end{array} 
\right) {\mathcal U}^\dagger + 
 {A (x)}   \left(
\begin{array}{ccc}
1+ \varepsilon_{ee}  & \varepsilon_{e \mu}  & 
\varepsilon_{e \tau}  \\ {\varepsilon_{e\mu} }^ \star & 
\varepsilon_{\mu \mu} &   \varepsilon_{\mu \tau} \\ 
{\varepsilon_{e \tau}}^\star & {\varepsilon_{\mu \tau}}^\star 
& \varepsilon_{\tau \tau}\\
\end{array} 
\right) \right\}  \ ,
 \eea 
where $A (x)=2 E \sqrt{2} G_F n_e (x)$ is the standard CC potential due to
the coherent forward scattering of neutrinos and $n_e$ is the electron
number density.  The three flavour neutrino mixing matrix ${\mathcal
  U}$ [$\equiv {\cal U}_{23} \, {\cal W}_{13} \,{\cal U}_{12}$ with
  ${\cal W}_{13} = {\cal U}_\delta~ {\cal U}_{13}~ {\cal
    U}_\delta^\dagger$ and ${\cal U}_\delta = {\mathrm{diag}}
  \{1,1,\exp{(i \delta)}\}$] is characterized by three angles and a
single (Dirac) phase and, in the standard PMNS parameterisation, we
have
\bea
{\mathcal U}^{} &=& \left(
\begin{array}{ccc}
1   & 0 & 0 \\  0 & c_{23}  & s_{23}   \\ 
 0 & -s_{23} & c_{23} \\
\end{array} 
\right)   
  \left(
\begin{array}{ccc}
c_{13}  &  0 &  s_{13} e^{- i \delta}\\ 0 & 1   &  0 \\ 
-s_{13} e^{i \delta} & 0 & c_{13} \\
\end{array} 
\right)  \left(
\begin{array}{ccc}
c_{12}  & s_{12} & 0 \\ 
-s_{12} & c_{12} &  0 \\ 0 &  0 & 1  \\ 
\end{array} 
\right)  \ ,
 \eea 
where $s_{ij}=\sin {\theta_{ij}}, c_{ij}=\cos \theta_{ij}$.   While, in addition, two Majorana phases are also
  possible, these are ignored as they play no role in neutrino
oscillations.  This particular 
 parameterisation  along with the fact of  ${\mathcal H}^{}_{\mathrm{SI}}$
commuting with ${\mathcal U}_{23}$, allows for a simplification. 
 Going over to the basis, $\tilde \nu = {\mathcal U}_{23}^\dagger
~\nu$,  we have $\tilde {\mathcal H} = {\mathcal U}_{23}^\dagger~ {\mathcal
  H} ~{\mathcal U}_{23} $
and\cite{Kikuchi:2008vq} 
\bea 
\label{htilde}
\tilde {\mathcal
H}^{}_{\mathrm{}} &=&
\lambda \Bigg[ \left(
\begin{array}{ccc}
r_A + s_{13}^2  & 0  & c_{13} s_{13} e^{-i \delta} \\  0&  0&0   \\ 
c_{13} s_{13} e^{i \delta} & 0  & c_{13}^2  \\
\end{array} 
\right)    + r_\lambda
 \left(
\begin{array}{ccc}
s_{12}^2 c_{13}^2   &  c_{12} s_{12} c_{13}
 & - s_{12}^2 c_{13} s_{13} e^{-i \delta}
  \\
  c_{12} s_{12} c_{13} &  c_{12}^2 & -c_{12} s_{12} s_{13} e^{-i \delta}   \\ 
s_{12}^2 c_{13} s_{13} e^{i\delta } & -c_{12} s_{12} s_{13} e^{i \delta} & 
s_{12}^2 s_{13}^2 \\
\end{array} 
\right)\nonumber\\
&& \quad + \quad
 r_A  \left(
\begin{array}{ccc}
{\tilde \varepsilon}_{ee}   & 
{\tilde \varepsilon}_{e \mu} & 
{\tilde \varepsilon}_{e \tau}  \\ 
 {\tilde \varepsilon}_{e\mu}^{ \star} & 
 {\tilde \varepsilon}_{\mu \mu} &  
  {\tilde \varepsilon}_{\mu \tau}^{} \\ 
{\tilde \varepsilon}_{e \tau}^{ \star} &
 {\tilde\varepsilon}_{\mu \tau}^{ \star} & 
 {\tilde \varepsilon}_{\tau \tau}^{} \\
\end{array} 
\right) \Bigg] \ ,
 \eea 
where we have defined the ratios 
\begin{equation}
\lambda \equiv \frac{\delta m^2_{31}}{2 E}  \quad \quad ; \quad \quad
r_{\lambda} \equiv \frac{\delta m^2_{21}}{\delta m^2_{31}} \quad \quad ; 
\quad \quad r_{A} \equiv \frac{A (x)}{\delta m^2_{31}} \ .
\label{dimless}
\end{equation}
Once again, $\tilde {\mathcal H}_{NSI} = {\mathcal U}_{23}^\dagger~ {\mathcal H}_{NSI} ~{\mathcal U}_{23} $ and  the last term in Eq.~(\ref{htilde}) is 
\bea 
 \lambda r_A
 \left(
\begin{array}{ccc}
{ \varepsilon}_{ee}   & c_{23} \varepsilon_{e \mu}  - s_{23} \varepsilon_
{e \tau}  
 & s_{23} \varepsilon_{e\mu} - c_{23} \varepsilon_
{e \tau} 
  \\ 
 c_{23} \varepsilon_{e \mu}^\star  - s_{23} \varepsilon_
{e \tau}^\star   & 
 \varepsilon_{\mu\mu} c_{23}^2  + \varepsilon_{\tau\tau} s_{23}^2
 -  (\varepsilon_{\mu\tau} + \varepsilon_{\mu\tau}^\star ) c_{23} 
 s_{23}   &  
  { \varepsilon}_{\mu \tau}^{} c_{23}^2 -\varepsilon_{\mu\tau}^\star s_{23}^2 + (\varepsilon_{\mu\mu} - \varepsilon_{\tau\tau}) c_{23} s_{23} \\ 
 s_{23} \varepsilon_{e\mu}^\star - c_{23} \varepsilon_
{e \tau} ^\star & 
  { \varepsilon}_{\mu \tau}^{\star} c_{23}^2 -\varepsilon_{\mu\tau} s_{23}^2 + (\varepsilon_{\mu\mu} - \varepsilon_{\tau\tau}) c_{23} s_{23} 
 & \varepsilon_{\mu\mu} s_{23}^2  + \varepsilon_{\tau\tau} c_{23}^2
 +  (\varepsilon_{\mu\tau} + \varepsilon_{\mu\tau}^\star ) c_{23} 
 s_{23} 
\end{array} 
\right) 
\nonumber
\eea
where ${\varepsilon}_{\alpha \beta} \, (\equiv |\varepsilon _{\alpha \beta}|
\, e^{i \phi_{\alpha\beta}})$ are complex.  For atmospheric and long
baseline neutrinos, $\lambda L \simeq {\cal O} (1)$ holds and $r_A L
\sim {\cal O} (1)$ for a large range of 
{the $E$ and $L$ values considered here}. The small quantities
are $r_\lambda \simeq 0.03$ and $ \tilde \varepsilon_{\alpha
\beta}$. We decompose $\tilde {\cal H}$ into two parts :
$\tilde {\cal H} = \tilde {\mathcal H}_0 + \tilde {\cal H} _I$ such
that the zeroth order term $\tilde {\mathcal H}_0$ provides the
effective two flavour limit with $r_A \neq 0$ and $s_{13} \neq 0$ but
$r_\lambda = 0$, \ie,
\bea 
\label{tildeh0}
\tilde {\mathcal
H}^{}_{\mathrm{0}} &=& \lambda  \left(
\begin{array}{ccc}
r_A (x) +s_{13}^2  & 0  & c_{13} s_{13} e^{-i \delta} \\  0&  0&0   \\ 
c_{13} s_{13} e^{i \delta} & 0  & c_{13}^2  \\
\end{array} 
\right)   \ ,
\eea 
while $\tilde {\cal H} _I$ contains the other two terms (on the RHS of
Eq.~(\ref{htilde})) which represent corrections due to non-zero
$r_\lambda$ and the non-zero NSI parameters $\tilde\varepsilon_{\alpha
\beta}$ respectively. Upon neglecting terms like $ r_\lambda s_{13},
r_\lambda s_{13}^2$, we get  an approximate form
for $\tilde {\cal H}_I$, {viz.,}
\bea 
\label{htildeone}
\tilde {\mathcal
H}^{}_{\mathrm{I}} &\approx&
\lambda \Bigg[  r_\lambda
 \left(
\begin{array}{ccc}
s_{12}^2   &  c_{12} s_{12} 
 & 0
  \\
  c_{12} s_{12}  &  c_{12}^2 & 0  \\ 
0  & 0 & 
0 \\
\end{array} 
\right)
  + 
 r_A  \left(
\begin{array}{ccc}
{\tilde \varepsilon}_{ee}   & 
{\tilde \varepsilon}_{e \mu} & 
{\tilde \varepsilon}_{e \tau}  \\ 
 {\tilde \varepsilon}_{e\mu}^{ \star} & 
 {\tilde \varepsilon}_{\mu \mu} &  
  {\tilde \varepsilon}_{\mu \tau}^{} \\ 
{\tilde \varepsilon}_{e \tau}^{ \star} &
 {\tilde\varepsilon}_{\mu \tau}^{ \star} & 
 {\tilde \varepsilon}_{\tau \tau}^{} \\
\end{array} 
\right) \Bigg] \ .
 \eea
 In what follows, we use the perturbation method described in~\cite{Akhmedov:2004ny} to compute the oscillation  probabilities.  The exact oscillation probability is given by 
 \bea
 P_{\alpha \beta} &=& |S _{\beta \alpha} (x,x_0)|^2 \ ,
 \eea 
 where $S (x,x_0)$ is the evolution matrix defined through $|\nu (x)
\rangle = S (x,x_0) ~ | \nu (x_0) \rangle$ with $S (x_0,x_0) =
{\mathbb I}$ and satisfying the same \sch equation as $|\nu (x)
\rangle$.  It can, trivially, be seen to be given by $S (x,x_0) =
{\mathcal U}_{23} ~\tilde S (x,x_0) ~{\mathcal U}_{23}^\dagger$ where
$\tilde S (x,x_0)$ is independent of $\theta_{23}$.  We first evaluate
$\tilde S_{} (x,{x_0})$ using
   \bea \tilde S_{} (x,{x_0}) &=& \tilde S_0 (x,x_0) ~ \tilde S_1 (x,x_0) \ . 
\eea
  {Here,} $\tilde S_0 (x,x_0)$ and $\tilde S_1 (x,x_0)$ satisfy
   \bea
 i\frac{d}{dx} \tilde S_0 (x,x_0) &=& \tilde {\cal H}_0 (x) ~ \tilde S_0  (x,x_0)~; \quad \tilde S_0 (x_0,x_0) = {\mathbb I} \ ,
\nonumber \\ 
 i\frac{d}{dx} \tilde S_1 (x,x_0) &=& [\tilde S_0 (x,x_0)^{-1} ~ \tilde {\cal H}_I (x) ~\tilde S_0 (x,x_0) ] \tilde S_1 (x,x_0) \quad ; \quad
  \tilde S_1 (x_0,x_0) = {\mathbb I} \ .
 \eea
 where $\tilde {\cal H}_I$ is given by Eq.~\ref{htildeone}.  To the first order in the expansion parameter, we have
  \bea
 \tilde S (x,x_0) &\simeq & \tilde S_0 (x,x_0) - i \tilde S_0 (x,x_0) \int_{x_0}^{x} [\tilde S_0 (x',x_0)^{-1} ~ \tilde {\mathcal H}_I (x')  ~ \tilde S_0 (x',x_0) ] dx'~.
\label{pert}
 \eea 
Finally, the full evolution matrix $S (x,x_0)$ can be obtained by going back to the original basis from the tilde basis using 
  $S (x,x_0) = {\mathcal U}_{23} ~\tilde S (x,x_0) ~{\mathcal U}_{23}^\dagger $.
The oscillation probability for $\nue \to \numu$ can be obtained as 
\begin{eqnarray}
 P_{e\mu}^{{{NSI} }} &\simeq& 
 4 s_{13}^2 s_{23}^2 \, \left[\dfrac{ \sin^2 {(1-r_A )\lambda L/2}{}}{ (1-r_A)^2} \right] \nonumber\\
 && + \, 8  s_{13} s_{23} c_{23}
( |\varepsilon_{e \mu}| c_{23} c_{\chi} - |\varepsilon_{e \tau}| s_{23} c_{\omega} ) \, r_A \,\left[\dfrac{\sin {r_A \lambda L/2}}{r_A} ~
  \dfrac{ \sin {(1 - r_A) \lambda L/2}}{(1-r_A)} ~\cos \frac{ \lambda L}{2}
 \right] \nonumber\\
  &&
+ \, 8  s_{13} s_{23} c_{23}
( |\varepsilon_{e \mu}| c_{23} s_{\chi} - |\varepsilon_{e \tau}| s_{23} s_{\omega} )r_A \,\left[\dfrac{\sin {r_A \lambda L/2}}{r_A} ~
  \dfrac{\sin {(1 - r_A) \lambda L/2}}{(1-r_A) } ~\sin \frac{ \lambda L}{2}
 \right]\nonumber\\
 && + \, 8  s_{13} s_{23}^2 
( |\varepsilon_{e \mu}| s_{23} c_{\chi} + |\varepsilon_{e \tau}| c_{23} c_{\omega} )r_A \,\left[
 \dfrac{\sin^2 {(1 - r_A) \lambda L/2} }{(1-r_A)^2} 
 \right] \ ,
 \label{pem}
 \end{eqnarray}
where we have used $\tilde s_{13} \equiv \sin \tilde \theta_{13} =
s_{13}/(1-r_A)$ to the leading order in $s_{13}$, and
$\chi=\phi_{e\mu}+ \delta$, $\omega = \phi_{e\tau}+\delta$. Only the
parameters $\varepsilon_{e \mu} $ and $ \varepsilon_{e\tau}$ enter in the
leading order expression~\cite{Kopp:2007ne}, {as} terms such as
$r_\lambda \varepsilon_{\alpha \beta}$ have been neglected.  Let us
discuss the two limiting cases, $r_A \to 0$ and $r_A \to 1$.  When
$r_A \to 0$, we recover the vacuum limit (given by {the} first
term on the RHS of Eq.~(\ref{pem})). When $r_A \to 1$, we are close to
the resonance condition ($r_A=\cos 2\theta_{13}$ since $\theta_{13}$
is small) and the probability remains finite due to the $(1-r_A)$ and
$(1-r_A)^2$ terms in the denominator of Eq.~(\ref{pem}).
   
The survival probability for $\numu \to \numu$ 
is given by
\begin{eqnarray}
\label{pmm}
 P_{\mu\mu}^{{{NSI}}}  &\simeq& 1 - s^2_{2\times {23}} \left[ \sin^2 \frac{\lambda L}{2} \right] \nonumber\\
 && - ~
 |\varepsilon_{\mu\tau}| \cos \phi_{\mu\tau} s_{2 \times {23}} \left[ s^2 _{2 \times {23}} (r_A \lambda L) \sin {{\lambda L}{}} + 4  c^2_{2 \times {23}}  r_A \sin^2 \frac{ \lambda L}{2} 
 \right]\nonumber\\
 && +~ (|\varepsilon _{\mu\mu}| - |\varepsilon _{\tau\tau}|) s^2_{2 \times {23}} c_{2 \times {23}}\left[  \dfrac{r_A \lambda L}{2} \sin  {\lambda L}{} - 2  r_A \sin^2 \frac{\lambda L}{2} \right] \ , 
 \end{eqnarray}
where $s_{2 \times {23}} \equiv \sin 2 \theta_{23} $ and $c_{2 \times
  {23}} \equiv \cos 2 \theta_{23} $. Note that the NSI parameters involving the electron sector do not enter this channel and the survival probability depends only on the three
parameters $\varepsilon_{\mu\mu}, \varepsilon_{\mu\tau},
\varepsilon_{\tau\tau}$~\cite{Kikuchi:2008vq,Coloma:2011rq,Kopp:2007ne}.  
{Once again, the vacuum limit is recovered for $r_A \to 0$.} 
Of these three NSI parameters, 
$\varepsilon_{\mu\mu}$ is subject to the most stringent
constraint (Eq.~\ref{largensi}).  If we look at 
$P_{\mu\mu}^{SI}$, the phase factor results in minima of
probability for  $ \lambda L/2 = (2p+1)\pi/2$ (vacuum dip)
 and maxima for $ \lambda L/2 = p\pi $
(vacuum peak) where $p$ is any integer. 
The oscillation length for the NSI terms, though, is different, 
and this changes the positions of the peaks and dips.

In order to quantify the impact of NSI, it is useful to define a difference\footnote{The difference used in~\cite{Ohlsson:2013epa} has an overall sign compared to our definition.}  
\begin{equation}
\Delta P_{\alpha\beta} = P_{\alpha\beta}^{SI} - P_{\alpha\beta}^{NSI} \ , 
\end{equation}
where $P_{\alpha\beta}^{SI}$ is probability of transition assuming
standard interactions (\ie, with $\varepsilon_{\alpha \beta}$ being set to
zero in Eqs.~(\ref{pem}) and (\ref{pmm})) and $ P_{\alpha\beta}^{NSI}$
is the transition probability in presence of NSI parameters.  For the
different channels that are relevant to our study, the quantities
$\Delta P_{\alpha\beta}$ are given by
\begin{eqnarray}
 \Delta P_{e\mu}  &\simeq&  
  - \, 8  s_{13} s_{23} c_{23}
( |\varepsilon_{e \mu}| c_{23} c_{\chi} - |\varepsilon_{e \tau}| s_{23} c_{\omega} ) \, r_A \,\left[\dfrac{\sin {r_A \lambda L/2}}{r_A} ~
  \dfrac{ \sin {(1 - r_A) \lambda L/2}}{(1-r_A)} ~\cos \frac{ \lambda L}{2}
 \right] \nonumber\\
  &&
- \, 8  s_{13} s_{23} c_{23}
( |\varepsilon_{e \mu}| c_{23} s_{\chi} - |\varepsilon_{e \tau}| s_{23} s_{\omega} )r_A \,\left[\dfrac{\sin {r_A \lambda L/2}}{r_A} ~
  \dfrac{\sin {(1 - r_A) \lambda L/2}}{(1-r_A) } ~\sin \frac{ \lambda L}{2}
 \right]\nonumber\\
 && - \, 8  s_{13} s_{23}^2 
( |\varepsilon_{e \mu}| s_{23} c_{\chi} + |\varepsilon_{e \tau}| c_{23} c_{\omega} )r_A \,\left[
 \dfrac{\sin^2 {(1 - r_A) \lambda L/2} }{(1-r_A)^2} 
 \right] \ .
 \label{delta_pem}
  \end{eqnarray}
\begin{eqnarray}
 \Delta P_{\mu\mu}  &\simeq& 
  |\varepsilon_{\mu\tau}| \cos \phi_{\mu\tau} s_{2\times {23}} \left( s^2_{2\times {23}}(r_A \lambda L)   \sin \lambda L + 4 c^2_{2\times {23}} r_A \sin^2 \frac{\lambda L}{2}\right) \nonumber \\
  && -~ (|\varepsilon _{\mu\mu}| - |\varepsilon _{\tau\tau}|) s^2_{2 \times {23}} c_{2 \times {23}}\left[  \dfrac{r_A \lambda L}{2} \sin  {\lambda L}{} - 2  r_A \sin^2 \frac{\lambda L}{2} \right] \ .
  \label{delta_pmm}
    \end{eqnarray} 
For the case of anti-neutrinos, $A \to -A$ (which implies that $r_A
\to - r_A$) while $\lambda \to \lambda, r_\lambda \to
r_\lambda$. Similarly for IH, $\lambda \to -\lambda, r_\lambda \to -
r_\lambda, r_A \to -r_A$.

In the present work, for the sake of simplicity, {the} NSI
parameters are taken to be real ($\varepsilon_{\alpha\beta} =
\varepsilon_{\alpha\beta}^\star$) and also $\delta_{} = 0$.

\section{Neutrino oscillograms in presence of NSI :}
\label{sec:oscillograms}

{Within the SM,} for a given hierarchy (NH or IH) and best-fit
values of the oscillation parameters (as given in
Table~\ref{tab_osc_param_input}), the oscillation probability depends
on only two quantities : the neutrino energy $E$ and the zenith angle
of the direction of the neutrino, namely $\theta$, with the vertically
downward direction corresponding to $\theta = 0$. The oscillation
pattern can, then, be fully described by contours of equal oscillation
probabilities in the $E-\cos \theta$ plane. We use these neutrino
oscillograms of Earth to discuss the effect of neutrino--matter
interactions on the atmospheric neutrinos passing through the Earth
(see Refs.~\cite{Akhmedov:2006hb,Akhmedov:2008qt} for a more detailed
discussion of the general features of the SI oscillograms).

{\bf{$\nu_\mu \to \nu_\mu$ disappearance channel :}}

\begin{figure}[!t]
\centering
 \includegraphics[width=1.0\textwidth, keepaspectratio]{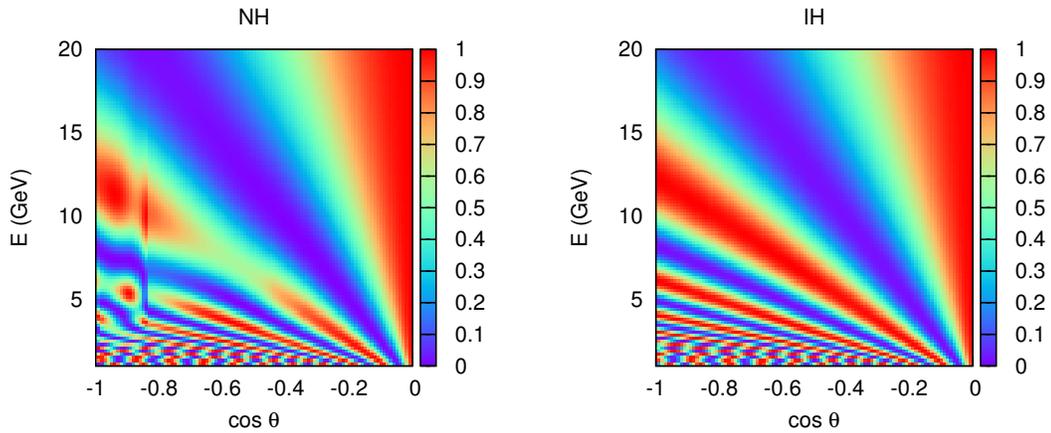}
\caption{
\label{pmm_si} Oscillograms of $P_{\mu\mu}$ for NH and IH with SI alone. 
}
\end{figure}
In Fig.~\ref{pmm_si}, we reproduce the neutrino oscillograms in the
$\nu_\mu \to \nu_\mu$ channel for the case of NH (left panel) and IH
(right panel) in the $E$-$\cos (\theta)$ plane.  As expected, the muon
neutrino disappearance probability experiences matter effects (MSW
effects {as well as parametric
resonances}) for the case of NH but not for the case of IH where it is
essentially given by the vacuum oscillation probability (which depends
on $\theta_{23}, \Delta m^2_{32}$). For SI in the $\bar\nu_\mu \to
\bar \nu_\mu$ channel, $A \to -A$ and the plots for NH and IH get
interchanged~\cite{Gandhi:2004bj}. In vacuum, the positions of the
peaks ($P_{\mu\mu} \simeq 1$) and dips ($P_{\mu\mu} \simeq 0$) can be
calculated from the first line on RHS in Eq.~(\ref{pmm}) as
\begin{equation}
{
(L/E)^{\rm{dip}} \simeq \frac{ (2p-1) \pi }{1.27 \times 2 \times \delta m^2_{31}}  ~{\textrm{km/GeV}} \quad {;} \quad (L/E)^{\rm{peak}} 
\simeq \frac{ k \pi }{1.27 \times \delta m^2_{31}}~{\textrm{km/GeV}}
}\end{equation}
where $p,k \in \mathbb{Z}^+$. The first dip and peak, then, are at
\begin{equation}
(L/E)^{\rm{dip}} \simeq 499  ~{\textrm{km/GeV}} \quad  {;} \quad (L/E)^{\rm{peak}} \simeq 998  ~{\textrm{km/GeV}}
\end{equation} 
which means that for a given $L$ (say $L=7000$ km or $\cos \theta = -0.549$), we can predict the
values of peak energy $E^{\rm{peak}}\sim 7$ GeV and dip energy
$E^{\rm{dip}} \sim 14$ GeV.  This can be seen clearly from the right
panel  of Fig.~\ref{pmm_si} which corresponds to the IH as the
probability in this case is  dominated by vacuum oscillations.
\begin{table}[!htb]
  \begin{center}
  \begin{tabular}{cccccc}
  \hline
 && &&&\\
\quad L \quad &  \quad $\cos \theta$ \quad & \quad $\rho_{\rm{avg}}$ \quad & \quad $E^{\rm{peak}}$ \quad & \quad $E^{\rm{dip}}$ \quad &\quad  $E_R$ \quad \\
 (km) & & ${\mathrm{(g/cc)}}$ & (GeV) & (GeV) & (GeV) \\
 \hline
 &&&&&\\
 3000 & $-$0.235 & 3.33 & 3.01 & 6.01 & 9.00 \\
 5000 & $-$0.392 & 3.68 & 5.01 & 10.02 & 8.14 \\
 7000 & $-$0.549 & 4.19 & 7.01 & 14.03 & 7.15 \\
 9000 & $-$0.706 & 4.56 & 9.02 & 18.04 & 6.57 \\
 11000 & $-$0.863 & 6.15 & 11.02 & 22.04 & 4.87 \\
 &&&&&
 \\
    \hline
  \end{tabular}
  \end{center}
  \caption{Values of  $E^{\rm{peak}}$ and $E^{\rm{dip}}$ in vacuum and $E_R$ for $P_{\mu\mu}$  as a function of $L$, $\cos \theta$ (for the choice of integers $p,k$ mentioned in the text).}
  \label{tab_epeakdip}
  \end{table}

%
The MSW matter effect can occur both in the mantle  region 
as well as the core~\cite{Wolfenstein:1977ue,Mikheev:1987qk}.  
The energy at which the MSW resonance takes place in the 13 sector is 
\begin{equation}
\rho E_R \simeq \dfrac{\delta m^2_{31}}{0.76 \times 10^{-4}} \times \cos 2\theta_{13} ~{\textrm{GeV g/cc}} \ .
\end{equation} 

Using the values of $\delta m^2_{31}$ and $\theta_{13}$ from
Table~\ref{tab_osc_param_input}, we get $E_R \sim 7.15~{\rm{GeV}}$ for
$\rho \simeq 4.19~{\textrm{g/cc}}$ which is  the average density for a neutrino traversing $\sim 7000$
  km through the earth\footnote{Note, though, that neutrinos of
      such energies but travelling a smaller path through the earth
      would also hit regions with $\rho \simeq 4.19~{\textrm{g/cc}}$
      and, thus, suffer resonant conversion.} to reach the
  detector.  As the neutrino path nears
the core, the energy at which the MSW resonance effects occur
decreases (see Table~\ref{tab_epeakdip}).  As discussed in
Ref.~\cite{Gandhi:2004bj}, when $E_R$ coincides with $E^{\rm{peak}}$
or $E^{\rm{dip}}$, one expects a large change in the probability. We
see this feature in the left plot of Fig.~\ref{pmm_si} around $E_R
\sim E^{\rm{peak}} \simeq 7$ GeV where the probability is reduced from
the peak value by almost $40\%$. (Note that $L = 7000$
  km ($\cos \theta \simeq -0.549$) implies that the neutrino has
  passed only through the crust and the mantle regions, without
  penetrating the core.).  Also the pattern in the left oscillogram
changes abruptly at a value of $\cos \theta_\nu = -0.84 $ demarcating
two regions : for $\cos \theta < -0.84 $, the neutrinos pass through
both mantle and core which allows for parametric effects while for
$\cos \theta > -0.84 $, the neutrinos cross only the mantle region
where only the usual MSW effects operate.  On the other hand, \sout{the}
parametric resonance occurs when neutrinos traversing the Earth pass
through layers of alternating density
(mantle-core-mantle)~\cite{Akhmedov:2006hb,Akhmedov:2008qt}.

Having described the case of SI, let us now address the impact of NSI
on neutrinos and antineutrinos traversing the Earth. To best illustrate
the features, we consider only one NSI parameter to be nonzero. In
the leading order expression  only two combinations of the
three NSI parameters ($\emt,\emm,\ett$)  appear. Let us discuss
{these in turn.}

\begin{figure}[!htb]
\centering
 \includegraphics[width=1.0\textwidth, keepaspectratio]{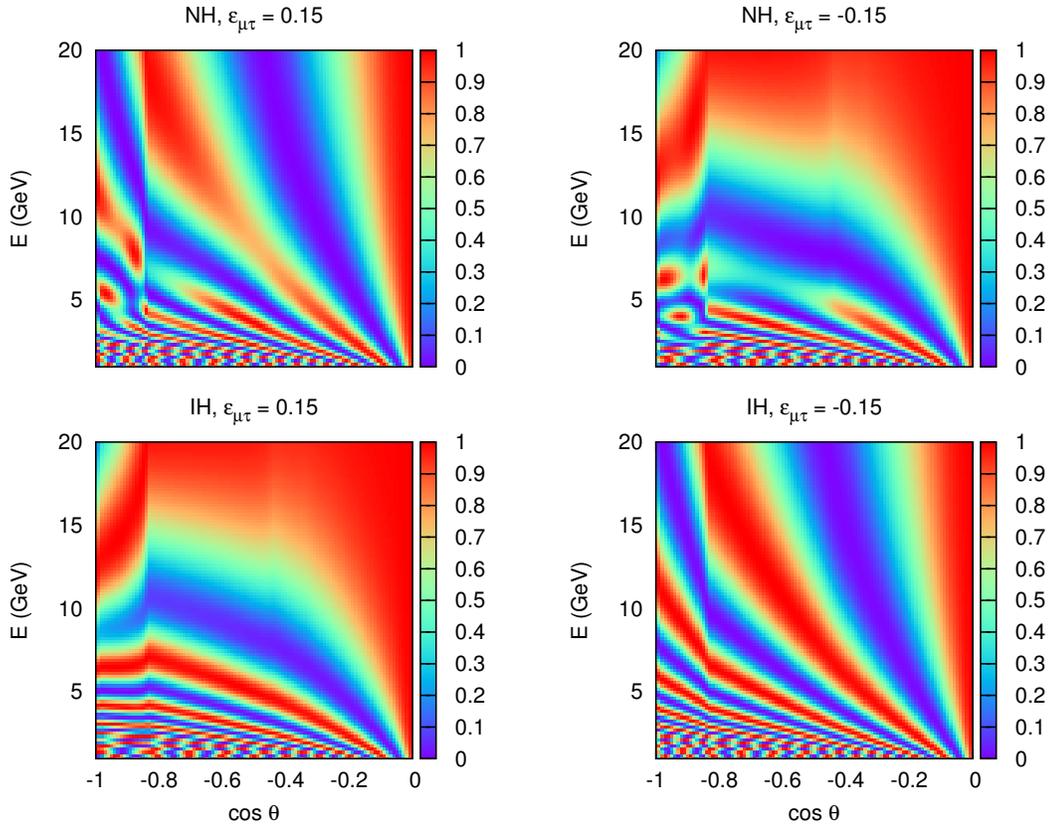}
\caption{
\label{pmm_nsi} Oscillograms of $P_{\mu\mu}$ for NH and IH with non-zero $\varepsilon_{\mu\tau}$.}
\end{figure}

\underline{(a) $\emt \neq 0; \varepsilon_{\mu\mu} = \varepsilon_{\tau\tau} =
  0$ :} In Fig.~\ref{pmm_nsi}, we show the corresponding
$P_{\mu\mu}^{NSI}$ for the case of NH (top row) and IH (bottom row)
and two specific values of the NSI parameter $\varepsilon_{\mu\tau}$
consistent with the current bounds.  Note that the case of NH and
$\varepsilon_{\mu\tau} > 0$ is grossly similar to the case of IH and
$\varepsilon_{\mu\tau} < 0$ (and, similarly, for NH and
$\varepsilon_{\mu\tau} < 0$ vs. IH and $\varepsilon_{\mu\tau} > 0$).  From
Eq.~(\ref{pmm}), we see that there are two terms proportional to
$\varepsilon _{\mu\tau}$, one where the oscillating function is $\sin
\lambda L$ with the other being $\sin^2 \lambda L/2$. Thus, the first
term can be positive or negative depending upon the value of the
phase, while the second term is always positive.  It is the interplay
of these two terms that leads to the features in these plots. The mass
hierarchy dependence comes from the first term since we have $r_A
\lambda L \sin (\lambda L)$ which changes sign when we go from NH to
IH. As noted earlier, near the vacuum dip $\lambda L = (2p+1) \pi/2$,
this term will be dominant. Consequently, for NH and
$\varepsilon_{\mu\tau} > 0$, the oscillatory pattern is a modification
of the standard one. For IH and $\varepsilon_{\mu\tau} > 0$, the term
proportional to $|\varepsilon_{\mu\tau} |$ will have a negative overall
sign and this leads to washout to a certain extent 
 of the oscillation pattern.

\begin{figure}[!htb]
\centering
 \includegraphics[width=1.0\textwidth, keepaspectratio]{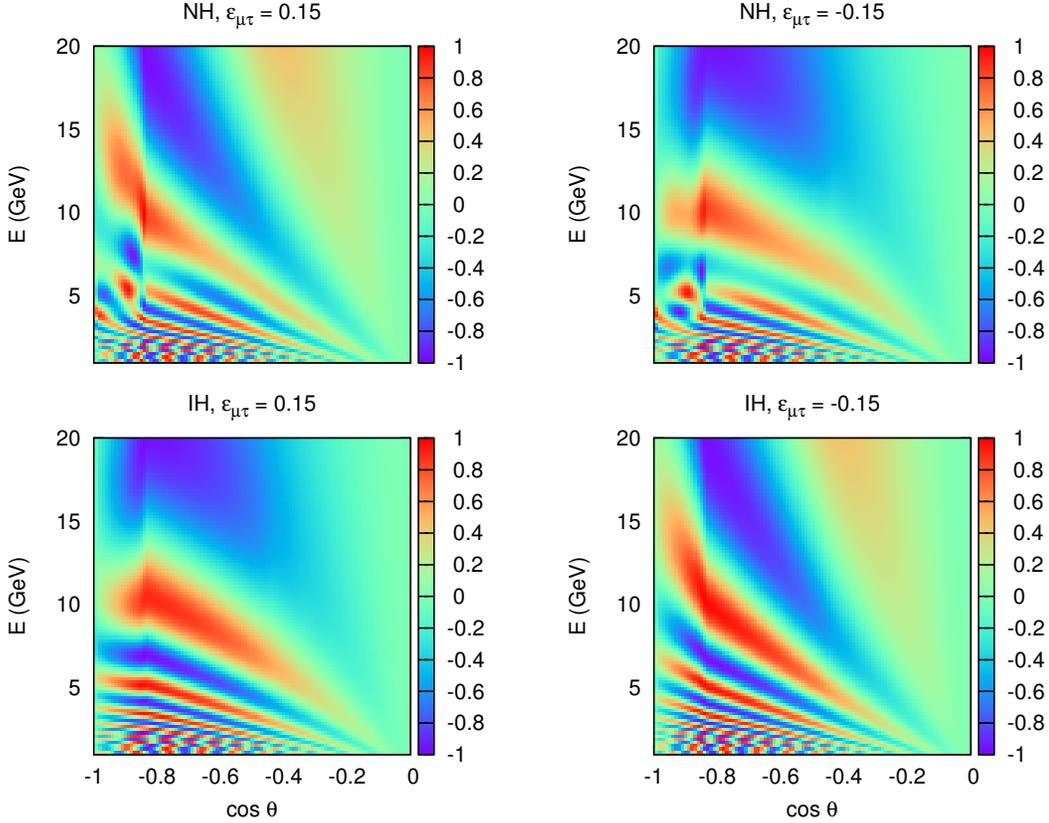}
\caption{
\label{deltapmm_nsi} 
Oscillograms of $\Delta P_{\mu\mu}$ with non-zero $\varepsilon_{\mu\tau}$. }
\end{figure}
The difference between SI and NSI contributions to the probability
$\Delta P_{\mu\mu}$ is shown in Fig.~\ref{deltapmm_nsi}. $|\Delta
P_{\mu\mu}|$ can be as large as 1 for regions in the core and in
mantle for some choice of $\varepsilon_{\mu\tau}$ and hierarchy.  We also
note large changes in probability (the regions where the difference is
large $\sim \pm 1$) along the diagonal line.

\begin{figure}[!htb]
\centering
 \includegraphics[width=1.0\textwidth, keepaspectratio]{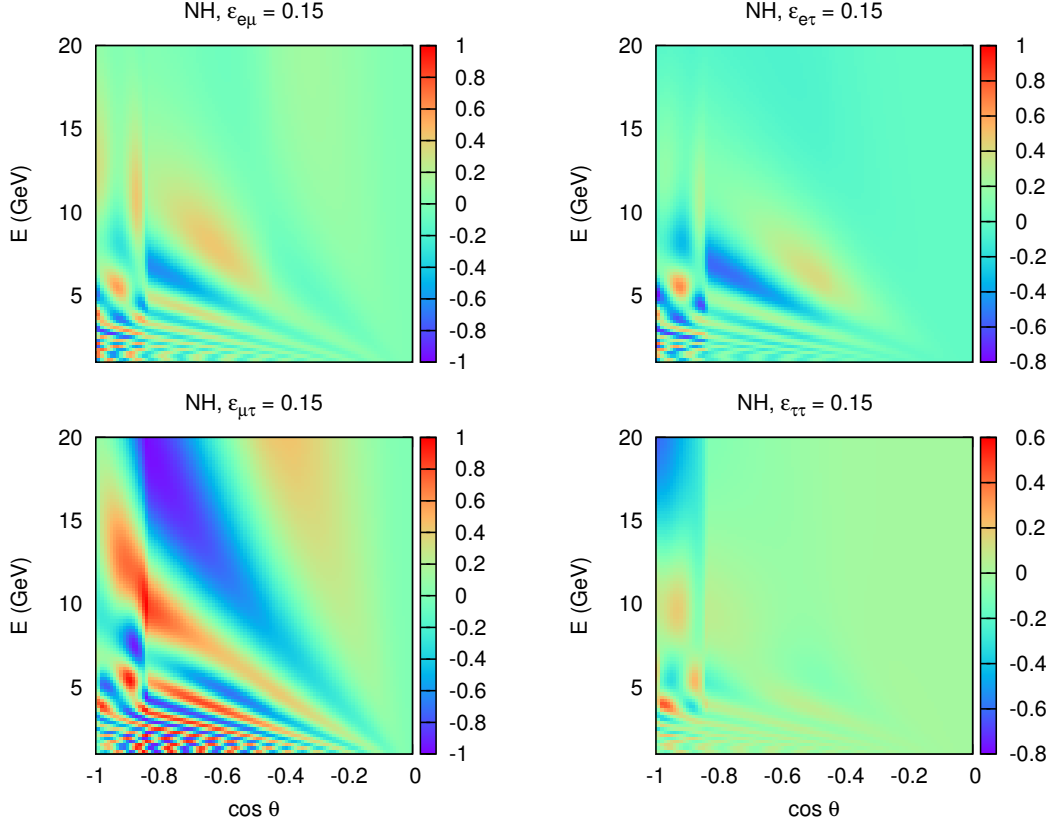}
\caption{
\label{figemutau_pmm}  Oscillogram pattern of $\Delta{P}_{{\mu}{\mu}}$ with non-zero  $\varepsilon_{e{\mu}},\varepsilon_{e\tau},\varepsilon_{{\mu}\tau}$ and $\varepsilon_{\tau\tau}$.}
\end{figure}

{{
\underline{(b) 
$(|\emm|-|\ett| ) \neq 0 ; \varepsilon_{\mu\tau} = 0$ :} This
case will correspond to the case of diagonal FP NSI parameter, 
$(|\varepsilon_{\mu\mu}| - |\varepsilon_{\tau\tau}|)$. As mentioned above, $|\varepsilon_{\mu\mu}|$ is
tightly constrained (see Eq.~(\ref{largensi})) while the bound on
$|\varepsilon_{\tau \tau}|$ is loose. If we choose 
$\varepsilon_{\tau\tau}=0.15$ (and $\varepsilon_{\mu\tau} = 0$), we see that effects 
due to $\varepsilon_{\tau\tau}$ in $\Delta P_{\mu\mu}$ are insignificant for most 
baselines except for a tiny region in the core (see Fig.~\ref{figemutau_pmm}). From 
Eq.~\ref{delta_pmm}, only the terms in second line contribute in this case and the minus sign between the two terms lowers the value of $\Delta P_{\mu\mu}$.}}

\underline{(c) Subdominant effects due to
  $\varepsilon_{e\mu},\varepsilon_{e\tau} \neq 0 $ :}  For the case of NH, we 
  compare the cases of non-zero
$\varepsilon_{\mu\tau}, \varepsilon_{e\mu}, \varepsilon_{e\tau}$ in
Fig.~\ref{figemutau_pmm}. From Eq.~(\ref{largensi}), we see that the
bounds for $\varepsilon_{e \mu}$ and $\varepsilon_{\mu \tau}$ are similar
($0.33$) while that on $\varepsilon_{e \tau}$ is rather loose ($3.0$).
It is seen that the other parameters involving the electron sector
play only a sub-dominant role in this channel. This can also be
understood from the fact that, in the expression for
  $P_{\mu\mu}^{NSI}$ (see Eq.~(\ref{pmm})), these terms appear only at
  the second order~\cite{Kikuchi:2008vq}.
\begin{figure}[!htb]
\centering
 \includegraphics[width=1.0\textwidth, keepaspectratio]{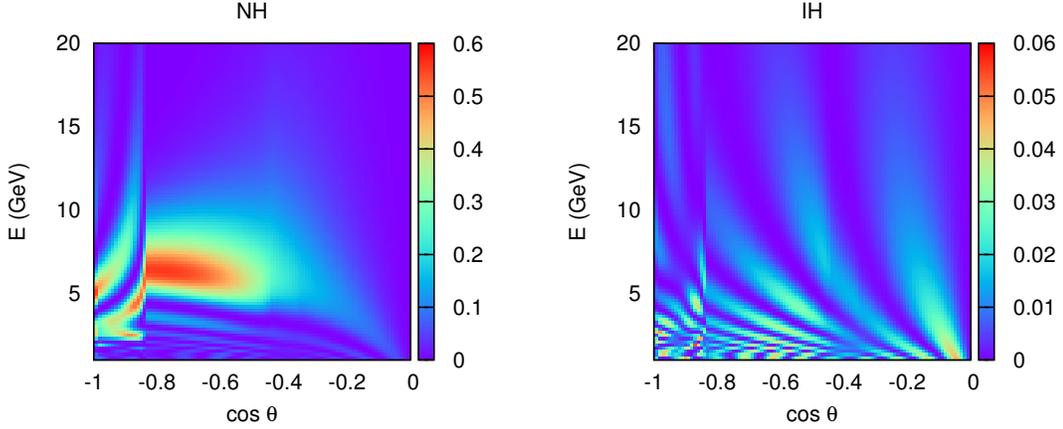}
\caption{
\label{pemu_si} Oscillogram for $P_{e\mu}$ for NH and IH for SI. 
}
\end{figure}
%

{\bf{$\nu_e \to \nu_\mu$ appearance channel : }}

In Fig.~\ref{pemu_si}, we have shown the standard neutrino
oscillograms in the $\nu_e \to \nu_\mu$ channel for the case of NH
(left panel) and IH (right panel) in the ($E$-$\cos \theta$) plane.
In this case, the probability is negligible in most parts of the
parameter space (especially for the case of IH). The $\nu_e \to
\nu_\mu$ appearance probability in matter differs from that in vacuum
in the leading order itself and also the position of peaks and
dips of the vacuum curves do not, in general, coincide with those in
presence of matter (unlike in the case of muon survival probability). In
order to analyse the $P_{e\mu}$ plots, let us look at the OMSD
expression~\cite{Gandhi:2004bj} (since our analytic expression is
valid to first order in $\theta_{13}$)
\begin{eqnarray}
 P_{e\mu}^{OMSD}  &=&  \sin^2 2 \tilde \theta_{13} \sin^2 {\theta_{23}}  \sin^2 \dfrac{\delta \tilde m_{31}^2 L}{4E}  \end{eqnarray}
 where
  \bea
 \sin 2 \tilde \theta_{13} &=& \sin 2 
 \theta_{13} \frac{\delta m^2_{31}}{\tilde {\delta m^2_{31}}}\nonumber\\
\delta \tilde m^2_{31} &\equiv& \sqrt{(\delta m_{31}^2 \cos 2\theta_{13} - A)^2 - (\delta m^2_{31} \sin 2\theta_{13})^2} 
 \eea 
The peak energy in matter will be given by~\cite{Gandhi:2004bj},
\begin{equation}
(L/E)^{\rm{peak}} \simeq \frac{ (2p-1) \pi }{1.27 \times 2 \times \tilde{ \delta m^2_{31}} } ~{\textrm{km/GeV}} 
\end{equation}
where $p \in \mathbb{Z}^+$.  One would
expect $P_{e\mu}$ to be large when the matter peak coincides with the
resonance energy, which gives $E_R \simeq 7$ GeV. However, the resonance
condition which implies that $ \sin 2 \tilde \theta_{13} = 1$ also
leads to $\tilde{\delta m^2_{31}} $ taking its minimum value at
resonance energy $\simeq \Delta m^2_{31} \sin 2\theta_{13} $.  Hence,
the probability becomes large when $\delta m^2_{31} \sin 2 \theta_{13}
L/4E \ge \pi/4$ is satisfied. This gives a value of $L=10,200$ km for
$\sin^2 2 \theta_{13} \simeq 0.1$~\cite{Gandhi:2004bj}.  Note
that the maximum value of $P_{e\mu}$ is given by the value of
$\sin^2\theta_{23} \sim 0.5$. The range of $E$ and $\cos \theta$ where
$P_{e\mu}$ is close to its maximal value due to MSW effect is given by
$E \in [5,7.5]$ GeV and $\cos \theta \in [-0.87, -0.5]$ in the mantle
region. In the core region, the MSW peak will occur at smaller energies
and the parametric resonance leads to large changes.
\begin{figure}[!htb]
\centering
 \includegraphics[width=1.0\textwidth, keepaspectratio]{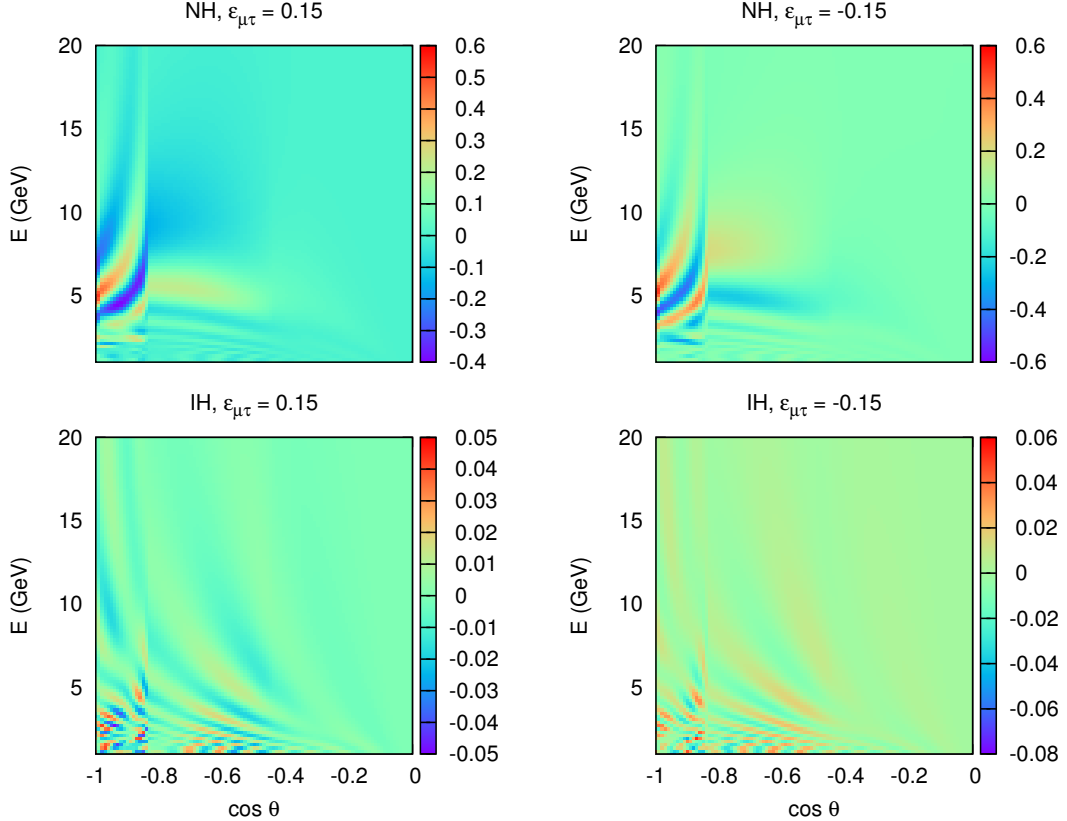}
\caption{
\label{deltapemu_nsi}
Oscillograms of $\Delta{P}_{e\mu}$ for  NSI parameter $\varepsilon_{\mu\tau}$. }
\end{figure}

Having described the case of SI, let us now address the impact of NSI
on neutrinos and antineutrinos traversing the Earth. In the leading
order expression for $P_{e\mu}^{NSI}$ (see Eq.~(\ref{pem})) there are
only two NSI parameters ($\eem,\eet$) that appear whereas
$\emt$ does not appear at all. We discuss them in turn.

 {\underline{(a) Subdominant effects due to $\emt \neq 0 $ :}} In
Fig.~\ref{deltapemu_nsi}, we show the effect of $\varepsilon_{\mu\tau}$
on the oscillograms of $\Delta{P}_{e\mu}$. Since the parameter
$\varepsilon_{\mu\tau}$ does not appear at all in the first order
expression (Eq.~(\ref{pem})), naturally its impact is expected to be
small. Consequently, $|\Delta P_{e\mu}| \neq 0$ only in very
tiny regions and can at best be as large as $0.3-0.4$.
 
\begin{figure}[!htb]
\centering
 \includegraphics[width=1.0\textwidth, keepaspectratio]{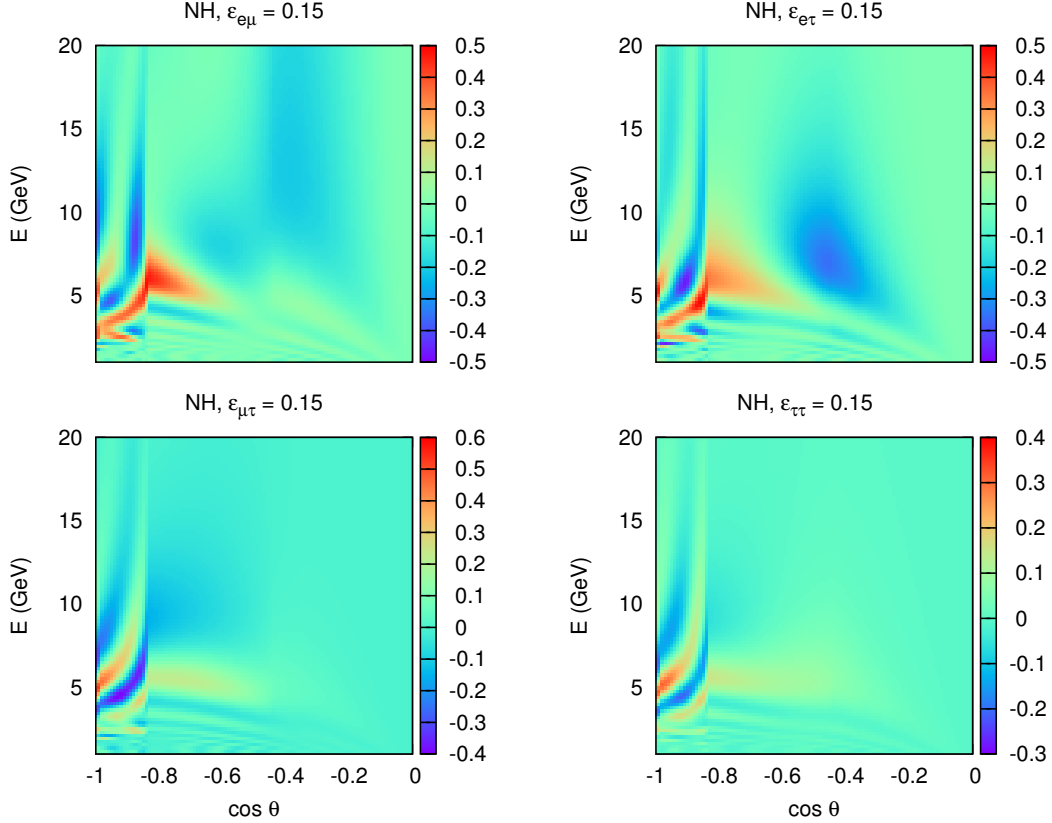}
\caption{
\label{figemutau_pem}  The effect of $\varepsilon_{e{\mu}}$,  $\varepsilon_{e\tau}$, $\varepsilon_{{\mu}\tau}$ and $\varepsilon_{\tau\tau}$ on the oscillogram of $\Delta{P}_{{e}{\mu}}$. 
}
\end{figure}

 {{
{\underline{(b) Comparison of effects due to $\eem \neq 0$, $\eet \neq
    0$, $\emt \neq 0$ and $\ett \neq 0$:}} In Fig.~\ref{figemutau_pem}, we compare 
    effects due to four NSI parameters for the case of NH,
allowing only one of them to be non-zero at a time. Since the
parameters $\varepsilon_{e\mu},\varepsilon_{e\tau}$ appear in the first
order expression (Eq.~(\ref{pem})), they naturally have a larger
impact as compared to the other two}} and, in the favourable
situation, $|\Delta P_{e\mu}| $ can be as large as $0.5$. This is to
be contrasted with $|\Delta P_{\mu\mu}| $ which could take values as
large as $1$ under favourable conditions. Also, if we look at
Eq.~(\ref{pem}), we note that $\eem$ and $\eet$ appear on equal
footing as far as $P_{e\mu}^{NSI}$ is concerned.

%


\section{{Simulating an experiment}}
\label{sec:analysis}

\subsection{Atmospheric events}

The neutrino and anti-neutrino CC events are obtained by  folding the 
incident neutrino fluxes with the appropriate  probabilities, relevant CC cross sections, the detector efficiency, resolution, mass and
the  exposure time. 

The $\mu^-$ event rate in a specific energy bin of width $\mathrm{dE}$ and the  angle bin of width ${\mathrm{d \Omega}}$ can be written as 
\beq 
\label{eq:muevent}
\mathrm{ \frac{d^2 N_{\mu}}{d \Omega \;dE} = \frac{1}{2\pi} ~\left[
\left(\frac{d^2 \Phi_\mu}{d \cos \theta \; dE}\right) P_{\mu\mu} +
\left(\frac{d^2 \Phi_e}{d \cos \theta \; dE}\right)
P_{e\mu}\right] ~\sigma_{CC}{ (\nu_\mu)} ~D_{eff}(\mu^-) }  \ .
\eeq

Here ${\mathrm{\Phi_{\mu,e}}}$ are the atmospheric fluxes ($\nu_\mu$
and ${\mathrm{\nu_e}}$), $\rm{\sigma_{CC}}$ is the total CC cross
section and $\rm{D_{eff}}$ is the detector efficiency.  {{
We have used the Honda atmospheric neutrino flux and cross-sections as given in~\cite{Gaisser:2002jj}.}} Similarly,
{the} $\mu^+$ event rate can be obtained {using the anti-neutrino
flux, probability and cross section, and the efficiency for $\mu^+$
(nominally, the same as for $\mu^-$)}. {Analogously, the
${\mathrm{e^-}}$ event rates would be given by}

\beq 
\label{eq:eevent}
\mathrm{ \frac{d^2 N_e}{d \Omega \;dE} = \frac{1}{2\pi}
~\left[\left( \frac{d^2 \Phi_{\mu}}{d \cos \theta \; dE}\right)
P_{\mu e} + \left(\frac{d^2 \Phi_e}{d \cos \theta \; dE}\right)
P_{e e} \right] ~\sigma_{CC}{ (\nu_e)} ~D_{eff}{ (e^-)} } \ ,
\eeq
with the ${\mathrm{e^+}}$ event rate being expressed in terms of
anti-neutrino fluxes, probabilities and cross sections {
as well $\text{D}_{\text{eff}} (e^+)$.}

In a realistic detector, the energy and angular resolution 
is not infinite, and to mimic this, we consider a Gaussian
 resolution
function, $R$. {For} the energy resolution function, we use
\beq 
\label{eq:esmear}
\mathrm{ R_{EN}(E_t,E_m) = \frac{1}{\sqrt{2\pi}\sigma} ~
\exp\left[-~\frac{(E_m - E_t)^2}{2 \sigma^2}\right]} \ , 
\eeq
where $\rm{E_m}$ and $\rm{E_t}$ denote the measured and true values of
energy respectively. The smearing width $\sigma$ is a function of
$\rm{E_t}$ {itself.} The functional form of $\sigma$ for ICAL and
\liar~detectors are given in Table.~\ref{tab:icalexptparams} and
\ref{tab:exptparams}.  Similarly, the angular smearing function is
given by
\beq
\label{eq:anglesmear}
\rm{ R_{\theta}(\Omega_t, \Omega_m) = N ~\exp \left[ -~ \frac{(\theta_t -
\theta_m)^2 + \sin^2 \theta_t ~(\phi_t - \phi_m)^2}{2 ~(\Delta
\theta)^2} \right] } \ , 
\eeq
where $N$ is a normalisation constant. 

The experimentally observable \numu\ event rates would, thus, be given by
\beq 
\label{eq:eventrate_m}
\rm{ \frac{d^2 N_{\mu}}{d \Omega_m ~dE_m} = \frac{1}{2\pi}
~\int \int dE_t~d\Omega_t~ R_{EN}(E_t,E_m)~
R_{\theta}(\Omega_t,\Omega_m)~\left[\Phi_{\mu}^d \; P_{\mu\mu} +
\Phi_{e}^{d} \;P_{e\mu} \right]~ \sigma_{CC}~ D_{eff} } \ ,
\eeq
and similarly for the \nue.  {Here}  we have denoted
$\rm{(d^2 \Phi/d \cos \theta \;dE)_{\mu,{e}} \equiv \Phi_{\mu,{e}}^d}$
\etc.  We limit the charged lepton phase space to $E_\ell \in [1,
10]$~GeV and $\cos\theta \in [-1.0, -0.1]$ which covers the incident
atmospheric neutrinos propagating through the earth. For effecting a
statistical analysis, we subdivide the energy ($\cos\theta$) range
into 9 (18) equal bins each.

It is worth noting at this stage that even if we incorporate the full
detector simulation for detectors such as the Iron Calorimeter
(ICAL) at INO or a generic Liquid Argon (LAr) one (such as in
Ref.~\cite{Chatterjee:2014vta}), the essential physics points of the
present work would not change. Since these studies are not yet
available for full-fledged reconstruction of neutrino energy and
angle using muons and hadrons, we {adopt} a simpler approach as
mentioned above.

\subsection{\chisq\ analysis}

We quantify the difference between the events with SI and NSI in terms
of a \chisq\ function.  For a fixed set of parameters, the latter is
calculated using the method of pulls, which allows us to take into
account the various statistical and systematic uncertainties (such as
those on the fluxes, cross sections \etc).

Let ${\mathrm{ N^{th}_{ij}(std) }}$ be the theoretical event rate for
the $\text{i-j}^{\text{th}}$ bin, as calculated with the standard
values for the inputs. Now, let us allow the ${\mathrm{ k^{th} }}$
input (known with an uncertainty ${\mathrm{ \sigma_k }}$) to deviate
from its standard value by an amount ${\mathrm{ \sigma_k \;\xi_k }}$.
If the relative uncertainties are not very large, the change in
${\mathrm{ N^{th}_{ij} }}$ can be expressed as a linear function of
the \emph{pull} variables ${\mathrm{ \xi_k }}$. In other words, the
value of ${\mathrm{ N^{th}_{ij} }}$ with the changed inputs is given
by
\begin{equation} 
\label{eqn:cij}
{\mathrm{ N^{th}_{ij} =  N^{th}_{ij}(std) + \sum^{npull}_{k=1}\;
c_{ij}^k \;\xi_k }} \ , 
\end{equation} 
where npull is the number of sources of uncertainty, which in
our case is 5. The systematic uncertainties are given in Table.~\ref{sys_uncern}.
 \begin{table}[!ht]

  \begin{center}
  \begin{tabular}{lc}
     
  \hline
  & \\
\quad  Uncertainty \quad  & \quad Value (in \%) \quad \\
 & \\
  \hline
  Flux Normalization & $20$ \\

  Tilt Factor & $5$\\
             
  Zenith angle dependence & $5$\\
             
  Cross section  & $10$\\

  Detector systematics & $5$ \\
                     
  \hline
  
  \end{tabular}
  \end{center}

  \caption{Different uncertainties used in our $\chi^{2}$ analysis.}
  \label{sys_uncern}
\end{table}
 {With these changed inputs, the goodness of fit 
is quantified in terms of a modified $\chi^2$ function defined as}

\begin{equation}  \label{eqn:chisq}
\mathrm{ {\chi^2(\xi_k)} = \sum_{i,j}\;
\frac{\left[~N_{ij}^{th}(std) \;+\; \sum^{npull}_{k=1}\;
c_{ij}^k\; \xi_k - N_{ij}^{ex}~\right]^2}{N_{ij}^{ex}} +
\sum^{npull}_{k=1}\; \xi_k^2  }
\end{equation}
where the additional term ${\rm{\xi_k^2}}$ is the penalty imposed
for moving {the} ${\mathrm{k^{th}}}$ input away from its standard value
by ${\rm{\sigma_k \;\xi_k}}$. The \chisq\ with pulls, which
includes the effects of all theoretical and systematic
uncertainties, is obtained by minimizing ${\rm{\chi^2(\xi_k)}}$,
given in Eq.~(\ref{eqn:chisq}), with respect to all the pulls ${\rm{\xi_k}}$,
viz.
\begin{equation}  \label{eqn:chisqmin}
{\mathrm{ \chi^2_{pull} = Min_{\xi_k}~ \left[~
\chi^2(\xi_k)~\right] }} \ .
\end{equation}

Note that ICAL magnetised detector will be able to distinguish muon
neutrinos and muon anti-neutrinos and hence the effective $\chi^2$ is
given by $\chi^2 _{\mu^-} + \chi^2 _{\mu^+}$. On the other hand, the
$\chi^2$ for the (unmagnetised) LIAR detector is $\chi^2 = \chi^2
_{\mu^- + \mu +} + \chi^2 _{e^- + e^+}$. Finally, we marginalize the
\chisq\ over the allowed range of the oscillation parameters as
mentioned in Table~\ref{tab_osc_param_input}.


\section{Event spectrum for the two detector types}
\label{sec:events}

We describe the details used for the two detector types (ICAL and LAr) used in our analysis :

{\sl{ICAL detector}}

{This} is a large magnetised iron detector and is
being planned for the INO experiment in South India. It consists of
151 layers of magnetized iron plates interleaved with Resistive Plate
Chambers (RPC) as active detector elements with a total mass of about
52  {kilotons}. Such a detector is capable of
detecting  muons (especially {for} GeV
energies) and identify their charge by virtue of {the}
magnetization. Additionally, {the} ICAL can detect hadronic
showers. The energy and angular resolution of muons and hadrons for
the ICAL have been obtained from {the} INO simulation code and
using that information the initial neutrino energy and angle can be
reconstructed. The detailed specifications are given in
Table~\ref{tab:icalexptparams}.

\begin{table}[htb!]
\centering
\begin{tabular}{l|l}
\hline
Energy Resolution ($\sigma (E)$) & $0.1 \sqrt{E}$\\

Angular Resolution ($\Delta \theta$)
  & $10^{\circ}$ \\

Detector efficiency ($\mathcal{E}$) & $85\%$ \\
\hline
\end{tabular}
\caption{\label{tab:icalexptparams} ICAL Detector parameters in the
  atmospheric neutrino experiment
  simulation~\cite{Chatterjee:2013qus}.}
\end{table}

\begin{figure}[!htb]
\centering
 \includegraphics[width=1.0\textwidth, keepaspectratio]{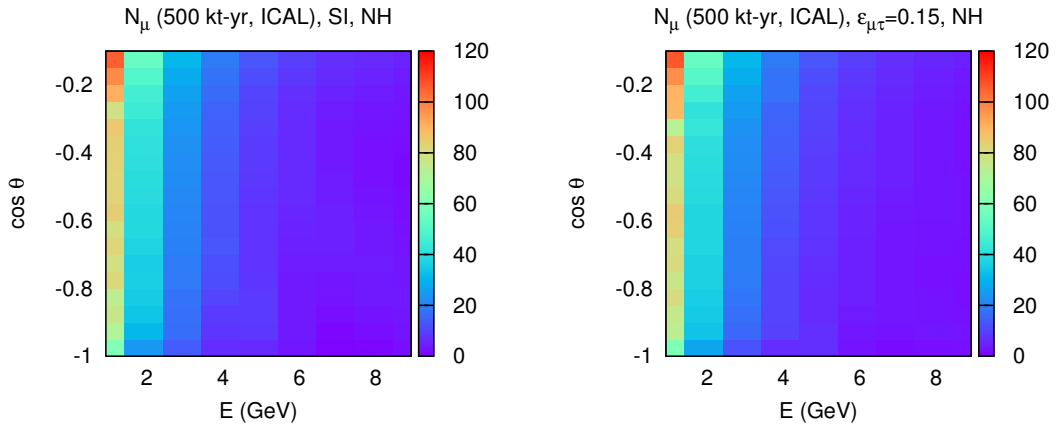}
\caption{
\label{fig9} $\nu_{\mu}$ events with SI and with NSI  for non-zero  $\varepsilon_{\mu\tau}$ (left). All the data are generated for 500 kT-yr of exposure for magnetized ICAL assuming NH as the true hierarchy. 
}
\end{figure}

\begin{figure}[!htb]
\centering
 \includegraphics[width=1.0\textwidth, keepaspectratio]{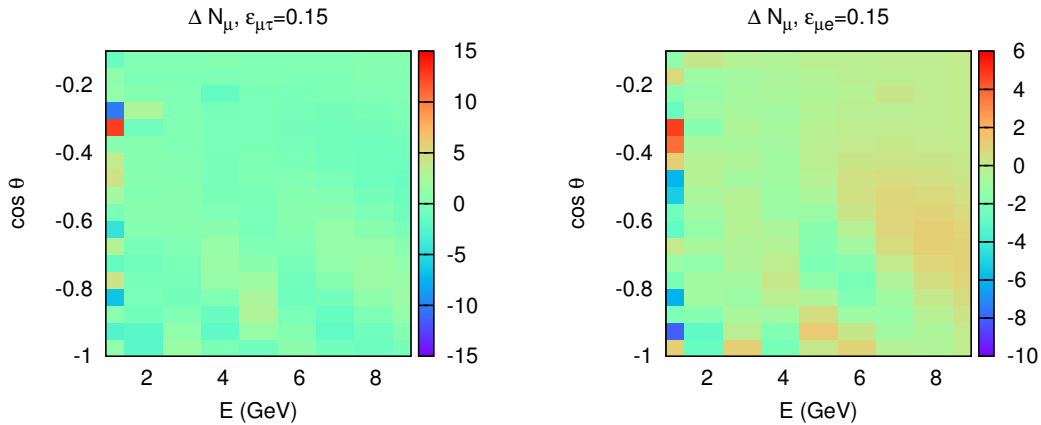}
\caption{The difference  with and without NSI  of $\nu_{\mu}$ (only) events for non-zero  $\varepsilon_{\mu\tau}$ (left) and $\varepsilon_{\mu{e}}$ (right). All the data are generated for 500 kT-yr of exposure for magnetized ICAL assuming NH as the true hierarchy. 
\label{fig10} 
}
\end{figure}

In Fig.~\ref{fig9}, the $\nu_\mu$ events are shown. At low energies,
the number of events is around $\sim 100$ for all the zenith angles
both for the case of SI and NSI ($\emt \neq 0$). The difference with
and without NSI of the $\nu_{\mu}$ events using parameters
$\varepsilon_{\mu\tau}$ and $\varepsilon_{\mu{e}}$ is shown in
Fig.~\ref{fig10}. For ICAL, it is evident that $\Delta N_{\mu} \simeq
\pm 10$ in some of the bins for $\varepsilon_{\mu\tau} \neq 0$ while for
$\varepsilon_{\mu{e}} \neq 0$, $\Delta N_{\mu} \sim \pm 4$.  This was
expected since the leading dependence was through
$\varepsilon_{\mu\tau}$, corroborates the probability level analysis.

\vskip 10pt
{\sl{LAr detector}}

A \liar~detector is capable of detecting not only muons but also
electrons, and has a very good angular resolution.  Since the detector
is unmagnetised, only the total events of a given flavour can be
measured.  For the proposed Long Baseline Neutrino Experiment {(which
is designated to operate with a beam and a baseline of 1300 km),} a 35
kt unmagnetized LAr detector {is to} be placed underground to study
atmospheric neutrinos along with the beam.  The specifications for the
\liar~detector are given in Table~\ref{tab:exptparams}. We shall
assume here a 10-yr operation period, or, equivalently, an effective
fiducial volume of 350 kt-yr.

\begin{table}[htb!]
\centering
\begin{tabular}{ l | l}
\hline
\multirow{2}{*}{Rapidity ($y$)} 
  & 0.45 for $\nu$\\
  & 0.30 for $\bar{\nu}$ \\

Energy Resolution ($\sigma (E)$) & $\sqrt{(0.01)^{2} + (0.15)^{2}/(yE_{}) + (0.03)^{2}}$\\

\multirow{2}{*}{Angular Resolution ($\Delta \theta$)}
  & $3.2^{\circ}$ for $\nu_{\mu}$	\\
  & $2.8^{\circ}$ for $\nu_{e}$ \\

Detector efficiency ($\mathcal{E}$) & $85\%$ \\
\hline
\end{tabular}
\caption{\label{tab:exptparams} The \liar~detector parameters used for  the atmospheric neutrino experiment simulation~\cite{Bueno:2007um,Barger:2012fx}.}
\end{table}
 
\begin{figure}[!htb]
\centering
 \includegraphics[width=1.0\textwidth, keepaspectratio]{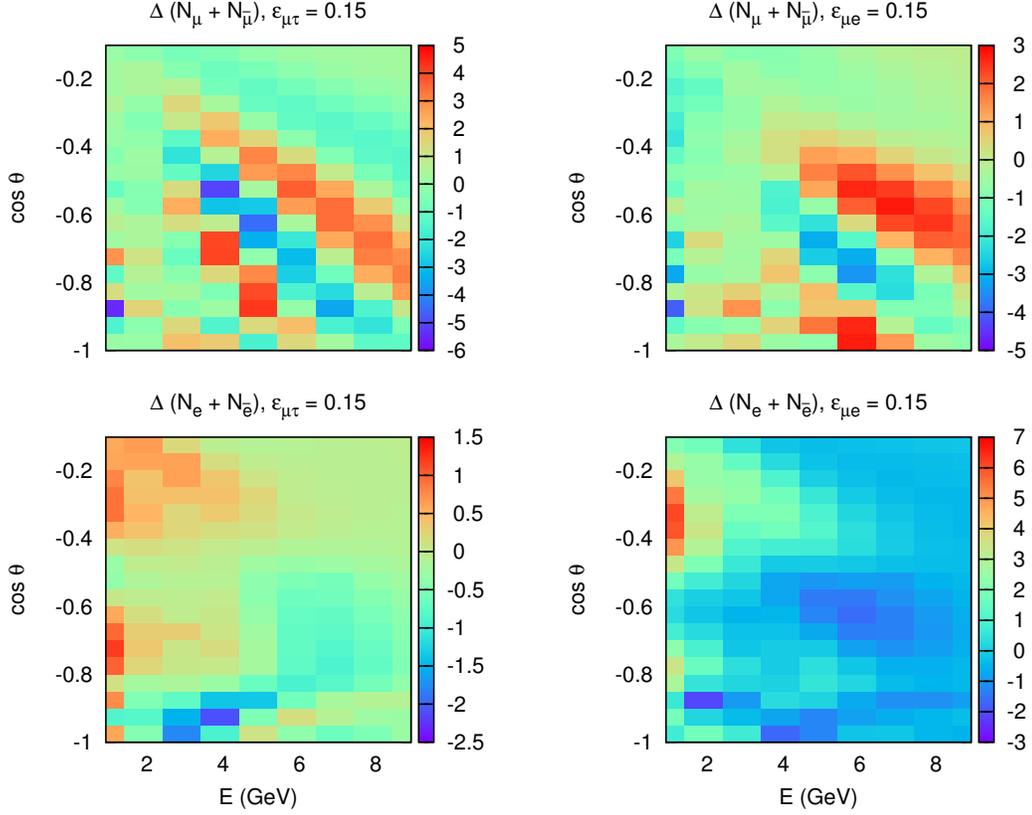}
\caption{The difference with and without NSI of the total muon neutrino events ($\nu_{\mu}$ + $\bar{\nu}_{\mu}$) and total electron neutrino events ($\nu_{e}$ + $\bar{\nu}_{e}$). All the data are generated for 350 kT-yr of exposure for unmagnetized \liar~detector assuming NH as the true hierarchy.
\label{fig12} 
}
\end{figure}

  The difference of the total muon ($\nu_{\mu}+\bar{\nu_{\mu}}$) and
  electron ($\nu_{e}+\bar{\nu_{e}}$) neutrino events with and without
  NSI are shown in Fig.~\ref{fig12}. \liar~ detector is complementary
  to ICAL as the impact of $\varepsilon_{\mu\tau}$ is less compared to
  $\varepsilon_{\mu{e}}$  for both muon and electron flavours.

\section{Results and Conclusions}

\label{sec:conclude}

\begin{figure}[!htb]
\centering
 \includegraphics[width=.8\textwidth]{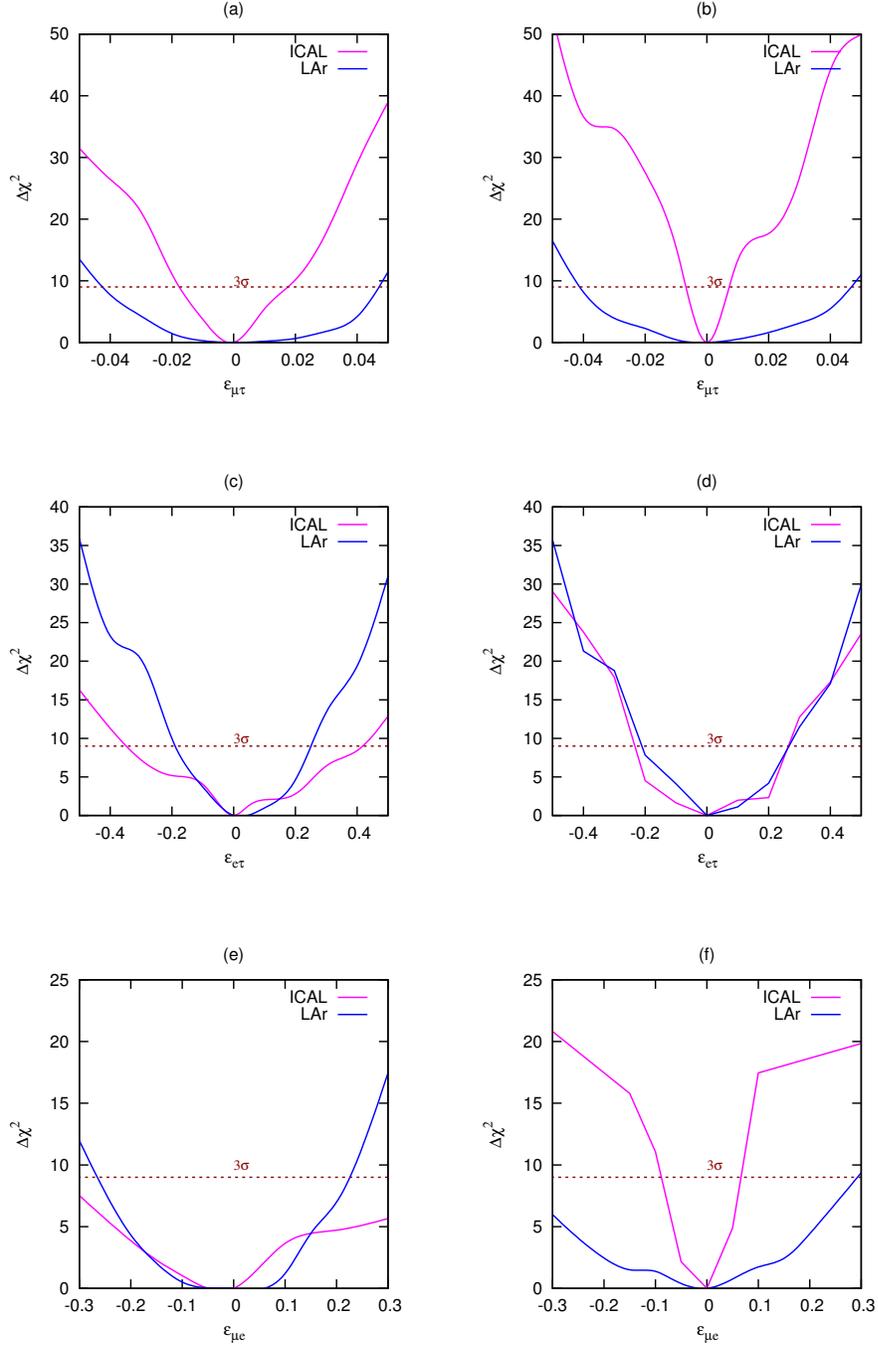}
\caption{$\Delta{\chi^{2}}$ vs $\varepsilon_{\mu\tau}$, $\varepsilon_{e\tau}$ and  $\varepsilon_{\mu e}$ for NH (left) and IH (right) for the two detector types.
\label{fig:chisq_mtau_et_me} 
}
\end{figure}

\begin{figure}[!htb]
\centering
 \includegraphics[width=.8\textwidth]{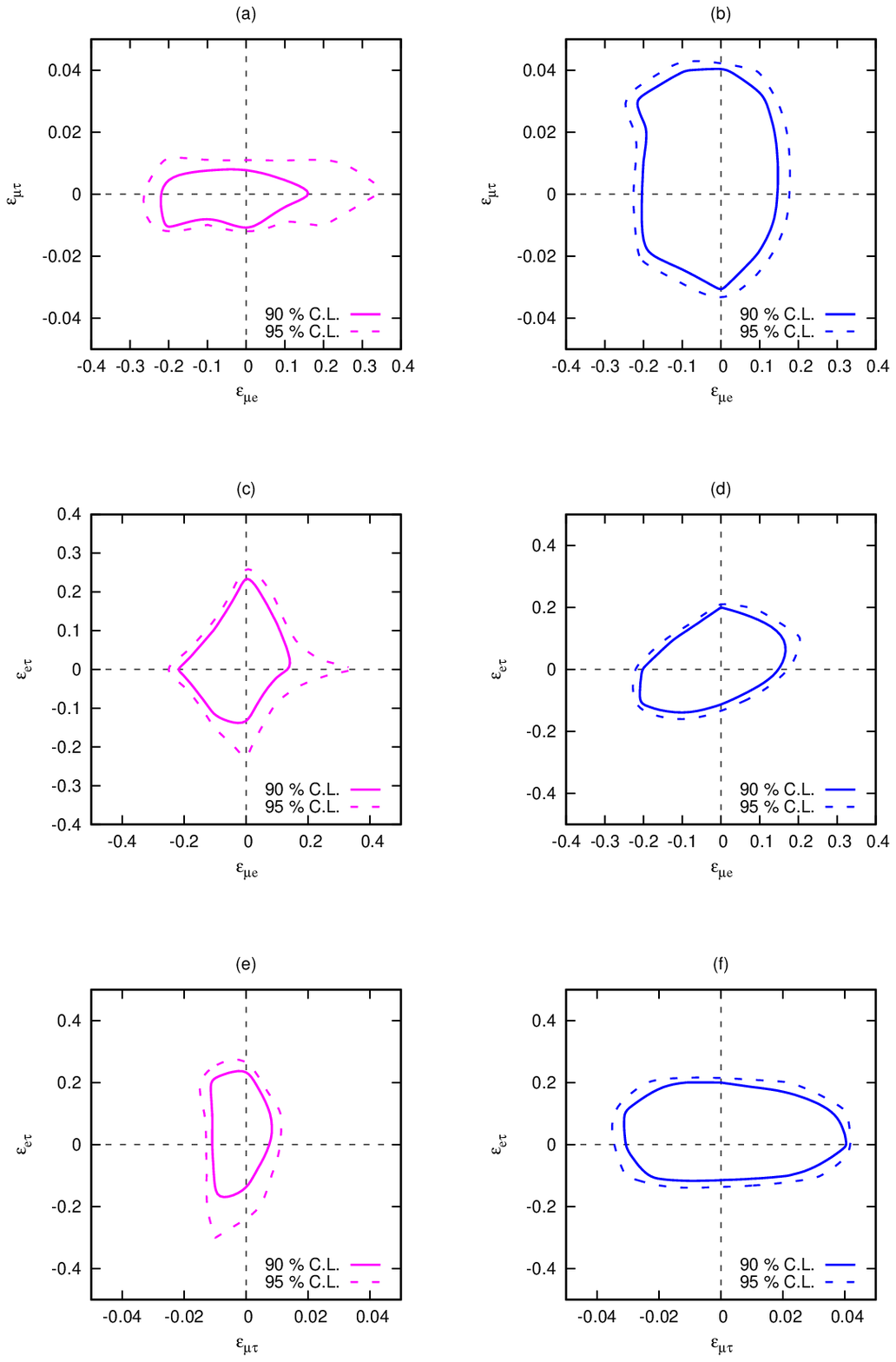}
\caption{Constraints on pairs of NSI parameters for different detector types. Left panel is for 500 kt-yr of fiducial volume of ICAL and right panel 
with 350 kt-yr unmagnetized LAr detector. NH is assumed to be true hierarchy.
\label{fig:contour_mtau_me} 
}
\end{figure}

\begin{figure}[!htb]
\centering
 \includegraphics[width=.8\textwidth]{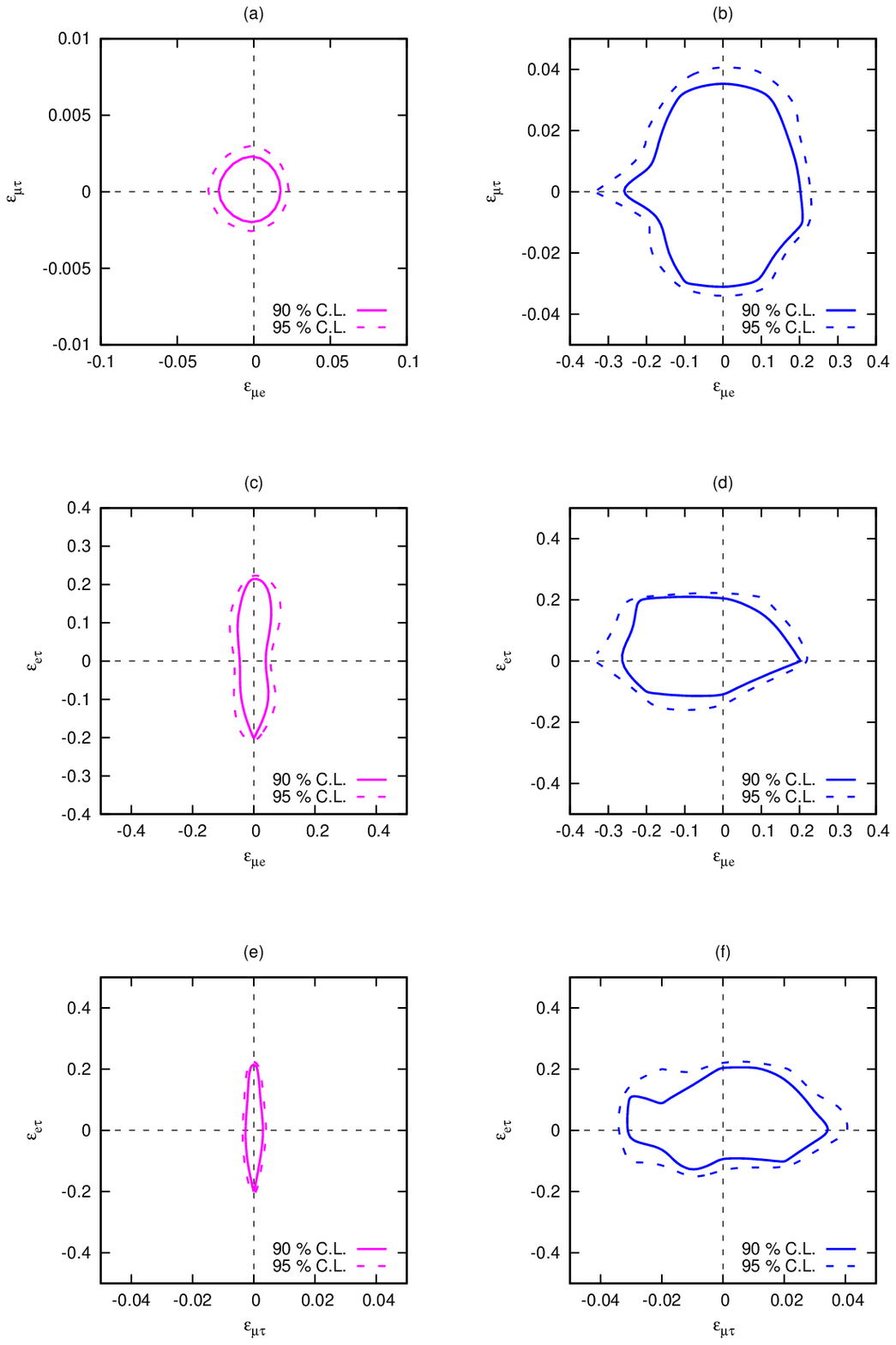}
\caption{Constraints on pairs of NSI parameters for different detector types. Left panel is for 500 kt-yr of fiducial volume of ICAL and right panel 
with 350 kt-yr unmagnetized LAr detector. IH is assumed to be true hierarchy.
\label{fig:contour_mtau_me_ih} 
}
\end{figure}







{We have considered the effect of NSI in the analysis of atmospheric neutrino oscillation 
experiments. To this end, we consider two detector types, {{\em viz.} a}
magnetised  iron one (the specifications being those
for the proposed ICAL) and  a generic unmagnetized
LAr detector and contrast the capabilities of these two detector types
for individual NSI parameters. With the constraints 
on $\varepsilon_{\mu\mu}$ already being very stringent~\cite{Escrihuela:2011cf}, we assume 
it is nonexistent, and concentrate on the others that such
detectors can be sensitive to, namely $\varepsilon_{\mu\tau}$,
$\varepsilon_{{\mu} e}$ and $\varepsilon_{e \tau}$.}

{In Fig~\ref{fig:chisq_mtau_et_me}, we show the variation of 
$\Delta{\chi^{2}}$ (for the individual detector choices)
with the aforementioned parameters. For the sake of simplicity, 
in each case, we vary only 
one of these parameters, assuming the other two to be vanishing identically.
For each plot, the horizontal
 dotted maroon line represents the $3\sigma$ CL bound that is expected 
to be reached.}

As far as $\varepsilon_{\mu\tau}$ is concerned, the dominant contribution comes from muon events and  the ICAL detector would perform better
than a \liar~detector both for the case of NH and IH (see
Figs.~11(a) and 11(b)). This is
 primarily due to the fact that the ICAL detector is magnetized which allows it to  distinguish between $\mu^+$ and $\mu^-$ events. In contrast, the unmagnetized \liar~detector can detect electron as well as muon events but is not able to identify charge of the leptons. 
The charge identification capability of ICAL allows us to add the two individual  $\Delta \chi^2$ corresponding to the $\mu^-$ and $\mu^+$ events respectively. For ICAL, the net $\Delta \chi^2$ is a sum of $\Delta \chi^2_{\mu^-}$ and $\Delta \chi^2_{\mu^+}$. For the \liar~detector, the net  $\Delta \chi^2$ comes from $\Delta \chi^2_{e^- + e^+}$ and $\Delta \chi^2_{\mu^- +\mu^+}$ and the lack of charge identification capability leads to poorer sensitivity.

As far as $\varepsilon_{e\tau}$  is concerned, the probabilities involving only the electron sector would play a role and the \liar~detector is expected to perform better than the ICAL detector (see Figs.~11(c) and  11(d)).  For NH, comparing the two plots (Figs.~11(a) and 11(c)), we see analogous effect for the case of NH with a role reversal of the two detectors. But for IH, comparing the Figs.~11(b) and 11(d), we note that for $e-\tau$ sector, there is very little difference between the curves of the two detectors and this is due to the fact that it is not possible to isolate the $e^-$ and $e^+$ events (as the \liar~detector is not magnetized). Another intriguing aspect is that the $\Delta \chi^2$ plot is asymmetric in the case of IH for the parameter $\varepsilon_{e\tau}$.

A nonzero $\varepsilon_{\mu e}$ manifests its presence in each
of $P_{\mu\mu}$, $P_{\mu e}$, $P_{e\mu}$ and $P_{ee}$ with the first 
two being more relevant, given the initial fluxes. Naively, thus,
one would expect the \liar\ detector to do better than the ICAL one. Once
again, though, the charge resolution can be  crucial. For the NH case, 
the \liar\ does win, but only marginally, and that too for 
large $|\varepsilon_{\mu e}|$ (where the ability to detect electron events 
becomes important). On the other hand, for the IH case, the structure 
of the matter effects ensures that the ICAL does much better.

It is worthwhile to look at the results above through the prism
of mass hierarchy. For the inverted scenario, the ICAL, thus, does
significantly better than a \liar\ detector for each of
$\varepsilon_{\mu\tau}$ and $\varepsilon_{\mu e}$, while for $\varepsilon_{e
\tau}$ the two perform similarly.  The reason is not difficult to
fathom. With $P_{\mu\mu}$ playing a major role in each of the first
two cases, the charge distinguishability available at the ICAL rules
the day, while this advantage is lost in the third.  More
intriguingly, the significantly better performance of the \liar\ for
negative $\varepsilon_{e \tau}$ can be traced back to the facilitation 
of the matter effect.

On the other hand, for NH, ICAL (\liar) does better 
for $\varepsilon_{\mu\tau}$ ($\varepsilon_{e\tau}$) whereas the performance 
is largely similar for $\varepsilon_{\mu e}$. Note that, for $\varepsilon_{\mu\tau}$,
the superiority of the ICAL is now less pronounced than it is for the inverted 
case. This degradation of the ICAL sensitivity, as well the corresponding 
one for $\varepsilon_{e\tau}$ can be traced back to the oscillograms of Sec.~4.
As the plots of $\Delta P_{\mu \mu}$ and $\Delta P_{\mu e}$
(Fig.~\ref{deltapmm_nsi} and \ref{deltapemu_nsi}) show, 
there are larger regions in $E-\cos \theta$ parameter space in case
of IH than NH for $\Delta P_{\mu\mu}$. Quite the
opposite is seen for $\Delta P _{\mu e}$ - the regions with
large change in probability actually shrink for IH compared to
NH. But, since the contribution from $P_{\mu e}$ to $N_{\mu}$ is
suppressed by the electron to muon flux ratio for the atmospheric
neutrinos and also the maximum possible change is $\sim \pm 0.5$
(which is much smaller than $\sim \pm 1$ for $\Delta P_{\mu \mu}$),
this does not nullify the large changes induced due to $P_{\mu\mu}$. 
It is, thus, amply clear that the unravelling of NSI parameters
requires detectors with complementary properties.

Expectedly,  different experimental systematics can lead to a quantitative change on the NSI parameters. However we believe that the choices we have made in this work are realistic, and  we  expect that other choices will not qualitatively alter the  conclusions presented.
 
These observations are also reflected in
Fig.~\ref{fig:contour_mtau_me} and ~\ref{fig:contour_mtau_me_ih} where the constraints on pairs of NSI
parameters, $\varepsilon_{{\mu\tau}}$ - $\varepsilon_{{\mu}e}$,
$\varepsilon_{{e\tau}}$ - $\varepsilon_{{\mu}e}$ and $\varepsilon_{{e\tau}}$ -
$\varepsilon_{{\mu}\tau}$ are shown {{for the case of NH and IH}}. 
 The allowed values of pairs of NSI parameters imply that we can demarcate between SI and NSI for those
values at a given confidence level. For ICAL (\liar) detector, the solid magenta (blue) line corresponds
to $90\%$ C.L. while the dashed magenta (blue)  line corresponds to $95\%$ C.L. 
The results are also summarised in Table~\ref{contourresult} 
{{for the two detector types for NH and IH. For ICAL, the expected sensitivity  is 
 better in case of IH in comparison to NH for the parameters $\emt$ and $\eme$ unlike the case 
 in~\cite{Choubey:2015xha} where the sensitivities in the two cases are comparable.   This is due to the fact that we have not used priors on standard parameters. We assume that the other experiments will significantly reduce the error bars on the standard parameters by the time these future atmospheric neutrino experiments are operational.
 For the $\emt$, we note that the bounds for ICAL are comparable to the ones obtained in 
 ~\cite{Choubey:2015xha} while for $\eme$ and $\eet$ our bounds are roughly
  a factor of two higher than those obtained in~\cite{Choubey:2015xha}. }}

 \begin{table}[!ht]

  \begin{center}
  \begin{tabular}{llll}
     
  \hline
  & &&\\
\quad  ICAL  (NH) \quad  &  \quad ICAL (IH)\quad&\quad  \liar~ (NH) 
\quad  &  \quad \liar~ (IH)\quad  \\
 & &  &\\
  \hline

$-0.02 < \emt < 0.01 $  & $ -0.005 < \emt < 0.005  $ & 
$-0.03 < \emt < 0.04 $  & $   -0.03 < \emt < 0.035 $\\

 $ -0.21 < \eme < 0.15 $ & $-0.07 < \eme < 0.05 $ & 
 $-0.2 < \eme < 0.15 $  & $ -0.25 < \eme < 0.20$\\
             
  $-0.2 < \eet < 0.23 $ & $ -0.2 < \eet < 0.2 $ & 
  $-0.12 < \eet < 0.2 $  & $ -0.15 < \eet < 0.2 $ \\
                     
  \hline
  
  \end{tabular}
  \end{center}
\begin{center}
  \end{center}

  \caption{Comparison of sensitivities offered by the two detectors 
  {{for NH and IH at 90\% CL. We assume 500 kt-yr  in case of ICAL and  350 kt-yr in case of LAr detector (see  Fig.~\ref{fig:contour_mtau_me} and ~\ref{fig:contour_mtau_me_ih}).  }}}
  \label{contourresult}
\end{table}

\section*{Acknowledgments :}{We would like to thank Brajesh Choudhary 
and Silvia Pascoli for discussions during the initial stages of this
work and Tommy Ohlsson for helpful email correspondence.  We
acknowledge the use of HRI cluster facility to carry out computations
in this work. We thank Mehedi Masud for crucial help with the plots. 
 AC thanks the INO Collaboration, Atri Bhattacharya,
Sandhya Choubey, Amol Dighe and Pomita Ghoshal for useful
discussions. DC and PM acknowledge the European Union grant FP7 ITN
INVISIBLES (Marie Curie Actions, PITN-GA-2011-289442). DC also
acknowledges the grant SR/MF/PS-02/2013-DUB from the Dept. of Science
and Technology, India.  RG acknowledges support from Fermi National
Accelerator Laboratory in the form of an Intensity Frontier
fellowship. PM acknowledges support from German Academic Exchange
Service (DAAD) for her visit to DESY, Zeuthen during which a major
part of this work was carried out and support from University Grants Commission under the second phase of  University with Potential of  Excellence at JNU. 

\bibliographystyle{apsrev}
\bibliography{referencesnsi}

\begin{thebibliography}{10}
\expandafter\ifx\csname bibnamefont\endcsname\relax
  \def\bibnamefont#1{#1}\fi
\expandafter\ifx\csname bibfnamefont\endcsname\relax
  \def\bibfnamefont#1{#1}\fi
\expandafter\ifx\csname url\endcsname\relax
  \def\url#1{\texttt{#1}}\fi
\expandafter\ifx\csname urlprefix\endcsname\relax\def\urlprefix{URL }\fi
\providecommand{\bibinfo}[2]{#2}
\providecommand{\eprint}[2][]{\url{#2}}

\bibitem{Fukuda:1998mi}
\bibinfo{author}{\bibfnamefont{Y.}~\bibnamefont{Fukuda}} \emph{et~al.}
  (\bibinfo{collaboration}{Super-Kamiokande Collaboration}),
  \bibinfo{journal}{Phys.Rev.Lett.} \textbf{\bibinfo{volume}{81}},
  \bibinfo{pages}{1562} (\bibinfo{year}{1998}), \eprint{hep-ex/9807003}.

\bibitem{GonzalezGarcia:2012sz}
\bibinfo{author}{\bibfnamefont{M.}~\bibnamefont{Gonzalez-Garcia}},
  \bibinfo{author}{\bibfnamefont{M.}~\bibnamefont{Maltoni}},
  \bibinfo{author}{\bibfnamefont{J.}~\bibnamefont{Salvado}}, \bibnamefont{and}
  \bibinfo{author}{\bibfnamefont{T.}~\bibnamefont{Schwetz}},
  \bibinfo{journal}{JHEP} \textbf{\bibinfo{volume}{1212}}, \bibinfo{pages}{123}
  (\bibinfo{year}{2012}), \eprint{1209.3023}.

\bibitem{Capozzi:2013csa}
\bibinfo{author}{\bibfnamefont{F.}~\bibnamefont{Capozzi}},
  \bibinfo{author}{\bibfnamefont{G.}~\bibnamefont{Fogli}},
  \bibinfo{author}{\bibfnamefont{E.}~\bibnamefont{Lisi}},
  \bibinfo{author}{\bibfnamefont{A.}~\bibnamefont{Marrone}},
  \bibinfo{author}{\bibfnamefont{D.}~\bibnamefont{Montanino}}, \emph{et~al.},
  \bibinfo{journal}{Phys.Rev.} \textbf{\bibinfo{volume}{D89}},
  \bibinfo{pages}{093018} (\bibinfo{year}{2014}), \eprint{1312.2878}.

\bibitem{Forero:2014bxa}
\bibinfo{author}{\bibfnamefont{D.~V.} \bibnamefont{Forero}},
  \bibinfo{author}{\bibfnamefont{M.}~\bibnamefont{T\'ortola}},
  \bibnamefont{and} \bibinfo{author}{\bibfnamefont{J.~W.~F.}
  \bibnamefont{Valle}}, \bibinfo{journal}{Phys. Rev. D}
  \textbf{\bibinfo{volume}{90}}, \bibinfo{pages}{093006}
  (\bibinfo{year}{2014}),
  \urlprefix\url{http://link.aps.org/doi/10.1103/PhysRevD.90.093006}.

\bibitem{Ohlsson:2012kf}
\bibinfo{author}{\bibfnamefont{T.}~\bibnamefont{Ohlsson}},
  \bibinfo{journal}{Rept.Prog.Phys.} \textbf{\bibinfo{volume}{76}},
  \bibinfo{pages}{044201} (\bibinfo{year}{2013}), \eprint{1209.2710}.

\bibitem{Grossman:1995wx}
\bibinfo{author}{\bibfnamefont{Y.}~\bibnamefont{Grossman}},
  \bibinfo{journal}{Phys.Lett.} \textbf{\bibinfo{volume}{B359}},
  \bibinfo{pages}{141} (\bibinfo{year}{1995}), \eprint{hep-ph/9507344}.

\bibitem{PhysRevLett.82.3202}
\bibinfo{author}{\bibfnamefont{M.~C.} \bibnamefont{Gonzalez-Garcia}},
  \bibinfo{author}{\bibfnamefont{M.~M.} \bibnamefont{Guzzo}},
  \bibinfo{author}{\bibfnamefont{P.~I.} \bibnamefont{Krastev}},
  \bibinfo{author}{\bibfnamefont{H.}~\bibnamefont{Nunokawa}},
  \bibinfo{author}{\bibfnamefont{O.~L.~G.} \bibnamefont{Peres}},
  \bibinfo{author}{\bibfnamefont{V.}~\bibnamefont{Pleitez}},
  \bibinfo{author}{\bibfnamefont{J.~W.~F.} \bibnamefont{Valle}},
  \bibnamefont{and}
  \bibinfo{author}{\bibfnamefont{R.}~\bibnamefont{Zukanovich~Funchal}},
  \bibinfo{journal}{Phys. Rev. Lett.} \textbf{\bibinfo{volume}{82}},
  \bibinfo{pages}{3202} (\bibinfo{year}{1999}).

\bibitem{Pilaftsis:2005rv}
\bibinfo{author}{\bibfnamefont{A.}~\bibnamefont{Pilaftsis}} \bibnamefont{and}
  \bibinfo{author}{\bibfnamefont{T.~E.} \bibnamefont{Underwood}},
  \bibinfo{journal}{Phys.Rev.} \textbf{\bibinfo{volume}{D72}},
  \bibinfo{pages}{113001} (\bibinfo{year}{2005}), \eprint{hep-ph/0506107}.

\bibitem{Barbieri:1988fh}
\bibinfo{author}{\bibfnamefont{R.}~\bibnamefont{Barbieri}} \bibnamefont{and}
  \bibinfo{author}{\bibfnamefont{R.~N.} \bibnamefont{Mohapatra}},
  \bibinfo{journal}{Phys.Lett.} \textbf{\bibinfo{volume}{B218}},
  \bibinfo{pages}{225} (\bibinfo{year}{1989}).

\bibitem{Babu:1989wn}
\bibinfo{author}{\bibfnamefont{K.}~\bibnamefont{Babu}} \bibnamefont{and}
  \bibinfo{author}{\bibfnamefont{R.}~\bibnamefont{Mohapatra}},
  \bibinfo{journal}{Phys.Rev.Lett.} \textbf{\bibinfo{volume}{63}},
  \bibinfo{pages}{228} (\bibinfo{year}{1989}).

\bibitem{Choudhury:1989pw}
\bibinfo{author}{\bibfnamefont{D.}~\bibnamefont{Choudhury}} \bibnamefont{and}
  \bibinfo{author}{\bibfnamefont{U.}~\bibnamefont{Sarkar}},
  \bibinfo{journal}{Phys.Lett.} \textbf{\bibinfo{volume}{B235}},
  \bibinfo{pages}{113} (\bibinfo{year}{1990}).

\bibitem{Healey:2013vka}
\bibinfo{author}{\bibfnamefont{K.~J.} \bibnamefont{Healey}},
  \bibinfo{author}{\bibfnamefont{A.~A.} \bibnamefont{Petrov}},
  \bibnamefont{and} \bibinfo{author}{\bibfnamefont{D.}~\bibnamefont{Zhuridov}},
  \bibinfo{journal}{Phys.Rev.}
  \textbf{\bibinfo{volume}{D87}}(\bibinfo{number}{11}), \bibinfo{pages}{117301}
  (\bibinfo{year}{2013}), \eprint{1305.0584}.

\bibitem{Bhatt:2009wb}
\bibinfo{author}{\bibfnamefont{J.~R.} \bibnamefont{Bhatt}},
  \bibinfo{author}{\bibfnamefont{B.~R.} \bibnamefont{Desai}},
  \bibinfo{author}{\bibfnamefont{E.}~\bibnamefont{Ma}},
  \bibinfo{author}{\bibfnamefont{G.}~\bibnamefont{Rajasekaran}},
  \bibnamefont{and} \bibinfo{author}{\bibfnamefont{U.}~\bibnamefont{Sarkar}},
  \bibinfo{journal}{Phys.Lett.} \textbf{\bibinfo{volume}{B687}},
  \bibinfo{pages}{75} (\bibinfo{year}{2010}), \eprint{0911.5012}.

\bibitem{Wetterich:2014gaa}
\bibinfo{author}{\bibfnamefont{C.}~\bibnamefont{Wetterich}},
  \bibinfo{journal}{Nuclear Physics B} \textbf{\bibinfo{volume}{897}},
  \bibinfo{pages}{111 } (\bibinfo{year}{2015}), ISSN \bibinfo{issn}{0550-3213},
  \urlprefix\url{http://www.sciencedirect.com/science/article/pii/S0550321315001790}.

\bibitem{Fornengo:2001pm}
\bibinfo{author}{\bibfnamefont{N.}~\bibnamefont{Fornengo}},
  \bibinfo{author}{\bibfnamefont{M.}~\bibnamefont{Maltoni}},
  \bibinfo{author}{\bibfnamefont{R.}~\bibnamefont{Tomas}}, \bibnamefont{and}
  \bibinfo{author}{\bibfnamefont{J.}~\bibnamefont{Valle}},
  \bibinfo{journal}{Phys.Rev.} \textbf{\bibinfo{volume}{D65}},
  \bibinfo{pages}{013010} (\bibinfo{year}{2002}), \eprint{hep-ph/0108043}.

\bibitem{Huber:2001zw}
\bibinfo{author}{\bibfnamefont{P.}~\bibnamefont{Huber}} \bibnamefont{and}
  \bibinfo{author}{\bibfnamefont{J.}~\bibnamefont{Valle}},
  \bibinfo{journal}{Phys.Lett.} \textbf{\bibinfo{volume}{B523}},
  \bibinfo{pages}{151} (\bibinfo{year}{2001}), \eprint{hep-ph/0108193}.

\bibitem{GonzalezGarcia:2004wg}
\bibinfo{author}{\bibfnamefont{M.}~\bibnamefont{Gonzalez-Garcia}}
  \bibnamefont{and} \bibinfo{author}{\bibfnamefont{M.}~\bibnamefont{Maltoni}},
  \bibinfo{journal}{Phys.Rev.} \textbf{\bibinfo{volume}{D70}},
  \bibinfo{pages}{033010} (\bibinfo{year}{2004}), \eprint{hep-ph/0404085}.

\bibitem{GonzalezGarcia:2011my}
\bibinfo{author}{\bibfnamefont{M.}~\bibnamefont{Gonzalez-Garcia}},
  \bibinfo{author}{\bibfnamefont{M.}~\bibnamefont{Maltoni}}, \bibnamefont{and}
  \bibinfo{author}{\bibfnamefont{J.}~\bibnamefont{Salvado}},
  \bibinfo{journal}{JHEP} \textbf{\bibinfo{volume}{1105}}, \bibinfo{pages}{075}
  (\bibinfo{year}{2011}), \eprint{1103.4365}.

\bibitem{Esmaili:2013fva}
\bibinfo{author}{\bibfnamefont{A.}~\bibnamefont{Esmaili}} \bibnamefont{and}
  \bibinfo{author}{\bibfnamefont{A.~Y.} \bibnamefont{Smirnov}},
  \bibinfo{journal}{JHEP} \textbf{\bibinfo{volume}{1306}}, \bibinfo{pages}{026}
  (\bibinfo{year}{2013}), \eprint{1304.1042}.

\bibitem{Datta2004356}
\bibinfo{author}{\bibfnamefont{A.}~\bibnamefont{Datta}},
  \bibinfo{author}{\bibfnamefont{R.}~\bibnamefont{Gandhi}},
  \bibinfo{author}{\bibfnamefont{P.}~\bibnamefont{Mehta}}, \bibnamefont{and}
  \bibinfo{author}{\bibfnamefont{S.~U.} \bibnamefont{Sankar}},
  \bibinfo{journal}{Physics Letters B}
  \textbf{\bibinfo{volume}{597}}(\bibinfo{number}{3–4}), \bibinfo{pages}{356
  } (\bibinfo{year}{2004}).

\bibitem{Chatterjee:2014oda}
\bibinfo{author}{\bibfnamefont{A.}~\bibnamefont{Chatterjee}},
  \bibinfo{author}{\bibfnamefont{R.}~\bibnamefont{Gandhi}}, \bibnamefont{and}
  \bibinfo{author}{\bibfnamefont{J.}~\bibnamefont{Singh}},
  \bibinfo{journal}{JHEP} \textbf{\bibinfo{volume}{1406}}, \bibinfo{pages}{045}
  (\bibinfo{year}{2014}), \eprint{1402.6265}.

\bibitem{Esmaili:2014ota}
\bibinfo{author}{\bibfnamefont{A.}~\bibnamefont{Esmaili}},
  \bibinfo{author}{\bibfnamefont{D.}~\bibnamefont{Gratieri}},
  \bibinfo{author}{\bibfnamefont{M.}~\bibnamefont{Guzzo}},
  \bibinfo{author}{\bibfnamefont{P.}~\bibnamefont{de~Holanda}},
  \bibinfo{author}{\bibfnamefont{O.}~\bibnamefont{Peres}}, \emph{et~al.},
  \bibinfo{journal}{Phys.Rev.} \textbf{\bibinfo{volume}{D89}},
  \bibinfo{pages}{113003} (\bibinfo{year}{2014}), \eprint{1404.3608}.

\bibitem{Esmaili:2014esa}
\bibinfo{author}{\bibfnamefont{A.}~\bibnamefont{Esmaili}},
  \bibinfo{author}{\bibfnamefont{O.}~\bibnamefont{Peres}}, \bibnamefont{and}
  \bibinfo{author}{\bibfnamefont{Z.}~\bibnamefont{Tabrizi}},
  \bibinfo{journal}{Journal of Cosmology and Astroparticle Physics}
  \textbf{\bibinfo{volume}{2014}}(\bibinfo{number}{12}), \bibinfo{pages}{002}
  (\bibinfo{year}{2014}),
  \urlprefix\url{http://stacks.iop.org/1475-7516/2014/i=12/a=002}.

\bibitem{Esmaili:2012nz}
\bibinfo{author}{\bibfnamefont{A.}~\bibnamefont{Esmaili}},
  \bibinfo{author}{\bibfnamefont{F.}~\bibnamefont{Halzen}}, \bibnamefont{and}
  \bibinfo{author}{\bibfnamefont{O.}~\bibnamefont{Peres}},
  \bibinfo{journal}{JCAP} \textbf{\bibinfo{volume}{1211}}, \bibinfo{pages}{041}
  (\bibinfo{year}{2012}), \eprint{1206.6903}.

\bibitem{Esmaili:2013cja}
\bibinfo{author}{\bibfnamefont{A.}~\bibnamefont{Esmaili}},
  \bibinfo{author}{\bibfnamefont{F.}~\bibnamefont{Halzen}}, \bibnamefont{and}
  \bibinfo{author}{\bibfnamefont{O.}~\bibnamefont{Peres}},
  \bibinfo{journal}{JCAP} \textbf{\bibinfo{volume}{1307}}, \bibinfo{pages}{048}
  (\bibinfo{year}{2013}), \eprint{1303.3294}.

\bibitem{Esmaili:2013vza}
\bibinfo{author}{\bibfnamefont{A.}~\bibnamefont{Esmaili}} \bibnamefont{and}
  \bibinfo{author}{\bibfnamefont{A.~Y.} \bibnamefont{Smirnov}},
  \bibinfo{journal}{JHEP} \textbf{\bibinfo{volume}{1312}}, \bibinfo{pages}{014}
  (\bibinfo{year}{2013}), \eprint{1307.6824}.

\bibitem{Datta:2000ci}
\bibinfo{author}{\bibfnamefont{A.}~\bibnamefont{Datta}},
  \bibinfo{author}{\bibfnamefont{R.}~\bibnamefont{Gandhi}},
  \bibinfo{author}{\bibfnamefont{B.}~\bibnamefont{Mukhopadhyaya}},
  \bibnamefont{and} \bibinfo{author}{\bibfnamefont{P.}~\bibnamefont{Mehta}},
  \bibinfo{journal}{Phys.Rev.} \textbf{\bibinfo{volume}{D64}},
  \bibinfo{pages}{015011} (\bibinfo{year}{2001}), \eprint{hep-ph/0011375}.

\bibitem{Mehta:2001na}
\bibinfo{author}{\bibfnamefont{P.}~\bibnamefont{Mehta}},
  \bibinfo{author}{\bibfnamefont{S.}~\bibnamefont{Dutta}}, \bibnamefont{and}
  \bibinfo{author}{\bibfnamefont{A.}~\bibnamefont{Goyal}},
  \bibinfo{journal}{Phys.Lett.} \textbf{\bibinfo{volume}{B535}},
  \bibinfo{pages}{219} (\bibinfo{year}{2002}), \eprint{hep-ph/0107214}.

\bibitem{Biggio:2009nt}
\bibinfo{author}{\bibfnamefont{C.}~\bibnamefont{Biggio}},
  \bibinfo{author}{\bibfnamefont{M.}~\bibnamefont{Blennow}}, \bibnamefont{and}
  \bibinfo{author}{\bibfnamefont{E.}~\bibnamefont{Fernandez-Martinez}},
  \bibinfo{journal}{JHEP} \textbf{\bibinfo{volume}{0908}}, \bibinfo{pages}{090}
  (\bibinfo{year}{2009}), \eprint{0907.0097}.

\bibitem{Davidson:2003ha}
\bibinfo{author}{\bibfnamefont{S.}~\bibnamefont{Davidson}},
  \bibinfo{author}{\bibfnamefont{C.}~\bibnamefont{Pena-Garay}},
  \bibinfo{author}{\bibfnamefont{N.}~\bibnamefont{Rius}}, \bibnamefont{and}
  \bibinfo{author}{\bibfnamefont{A.}~\bibnamefont{Santamaria}},
  \bibinfo{journal}{JHEP} \textbf{\bibinfo{volume}{0303}}, \bibinfo{pages}{011}
  (\bibinfo{year}{2003}), \eprint{hep-ph/0302093}.

\bibitem{Escrihuela:2011cf}
\bibinfo{author}{\bibfnamefont{F.}~\bibnamefont{Escrihuela}},
  \bibinfo{author}{\bibfnamefont{M.}~\bibnamefont{Tortola}},
  \bibinfo{author}{\bibfnamefont{J.}~\bibnamefont{Valle}}, \bibnamefont{and}
  \bibinfo{author}{\bibfnamefont{O.}~\bibnamefont{Miranda}},
  \bibinfo{journal}{Phys.Rev.} \textbf{\bibinfo{volume}{D83}},
  \bibinfo{pages}{093002} (\bibinfo{year}{2011}), \eprint{1103.1366}.

\bibitem{Mitsuka:2011ty}
\bibinfo{author}{\bibfnamefont{G.}~\bibnamefont{Mitsuka}} \emph{et~al.}
  (\bibinfo{collaboration}{Super-Kamiokande Collaboration}),
  \bibinfo{journal}{Phys.Rev.} \textbf{\bibinfo{volume}{D84}},
  \bibinfo{pages}{113008} (\bibinfo{year}{2011}), \eprint{1109.1889}.

\bibitem{Adamson:2013ovz}
\bibinfo{author}{\bibfnamefont{P.}~\bibnamefont{Adamson}} \emph{et~al.}
  (\bibinfo{collaboration}{MINOS Collaboration}), \bibinfo{journal}{Phys.Rev.}
  \textbf{\bibinfo{volume}{D88}}(\bibinfo{number}{7}), \bibinfo{pages}{072011}
  (\bibinfo{year}{2013}), \eprint{1303.5314}.

\bibitem{Kopp:2010qt}
\bibinfo{author}{\bibfnamefont{J.}~\bibnamefont{Kopp}},
  \bibinfo{author}{\bibfnamefont{P.~A.} \bibnamefont{Machado}},
  \bibnamefont{and} \bibinfo{author}{\bibfnamefont{S.~J.} \bibnamefont{Parke}},
  \bibinfo{journal}{Phys.Rev.} \textbf{\bibinfo{volume}{D82}},
  \bibinfo{pages}{113002} (\bibinfo{year}{2010}), \eprint{1009.0014}.

\bibitem{Ohlsson:2013epa}
\bibinfo{author}{\bibfnamefont{T.}~\bibnamefont{Ohlsson}},
  \bibinfo{author}{\bibfnamefont{H.}~\bibnamefont{Zhang}}, \bibnamefont{and}
  \bibinfo{author}{\bibfnamefont{S.}~\bibnamefont{Zhou}},
  \bibinfo{journal}{Phys.Rev.}
  \textbf{\bibinfo{volume}{D88}}(\bibinfo{number}{1}), \bibinfo{pages}{013001}
  (\bibinfo{year}{2013}), \eprint{1303.6130}.

\bibitem{Gonzalez-Garcia:2013usa}
\bibinfo{author}{\bibfnamefont{M.}~\bibnamefont{Gonzalez-Garcia}}
  \bibnamefont{and} \bibinfo{author}{\bibfnamefont{M.}~\bibnamefont{Maltoni}},
  \bibinfo{journal}{JHEP} \textbf{\bibinfo{volume}{1309}}, \bibinfo{pages}{152}
  (\bibinfo{year}{2013}), \eprint{1307.3092}.

\bibitem{Barger:1980tf}
\bibinfo{author}{\bibfnamefont{V.~D.} \bibnamefont{Barger}},
  \bibinfo{author}{\bibfnamefont{K.}~\bibnamefont{Whisnant}},
  \bibinfo{author}{\bibfnamefont{S.}~\bibnamefont{Pakvasa}}, \bibnamefont{and}
  \bibinfo{author}{\bibfnamefont{R.}~\bibnamefont{Phillips}},
  \bibinfo{journal}{Phys.Rev.} \textbf{\bibinfo{volume}{D22}},
  \bibinfo{pages}{2718} (\bibinfo{year}{1980}).

\bibitem{Cervera:2000kp}
\bibinfo{author}{\bibfnamefont{A.}~\bibnamefont{Cervera}},
  \bibinfo{author}{\bibfnamefont{A.}~\bibnamefont{Donini}},
  \bibinfo{author}{\bibfnamefont{M.}~\bibnamefont{Gavela}},
  \bibinfo{author}{\bibfnamefont{J.}~\bibnamefont{Gomez~Cadenas}},
  \bibinfo{author}{\bibfnamefont{P.}~\bibnamefont{Hernandez}}, \emph{et~al.},
  \bibinfo{journal}{Nucl.Phys.} \textbf{\bibinfo{volume}{B579}},
  \bibinfo{pages}{17} (\bibinfo{year}{2000}), \eprint{hep-ph/0002108}.

\bibitem{Gandhi:2004md}
\bibinfo{author}{\bibfnamefont{R.}~\bibnamefont{Gandhi}},
  \bibinfo{author}{\bibfnamefont{P.}~\bibnamefont{Ghoshal}},
  \bibinfo{author}{\bibfnamefont{S.}~\bibnamefont{Goswami}},
  \bibinfo{author}{\bibfnamefont{P.}~\bibnamefont{Mehta}}, \bibnamefont{and}
  \bibinfo{author}{\bibfnamefont{S.~U.} \bibnamefont{Sankar}},
  \bibinfo{journal}{Phys.Rev.Lett.} \textbf{\bibinfo{volume}{94}},
  \bibinfo{pages}{051801} (\bibinfo{year}{2005}), \eprint{hep-ph/0408361}.

\bibitem{Indumathi:2004kd}
\bibinfo{author}{\bibfnamefont{D.}~\bibnamefont{Indumathi}} \bibnamefont{and}
  \bibinfo{author}{\bibfnamefont{M.}~\bibnamefont{Murthy}},
  \bibinfo{journal}{Phys.Rev.} \textbf{\bibinfo{volume}{D71}},
  \bibinfo{pages}{013001} (\bibinfo{year}{2005}), \eprint{hep-ph/0407336}.

\bibitem{Choudhury:2004sv}
\bibinfo{author}{\bibfnamefont{D.}~\bibnamefont{Choudhury}} \bibnamefont{and}
  \bibinfo{author}{\bibfnamefont{A.}~\bibnamefont{Datta}},
  \bibinfo{journal}{JHEP} \textbf{\bibinfo{volume}{0507}}, \bibinfo{pages}{058}
  (\bibinfo{year}{2005}), \eprint{hep-ph/0410266}.

\bibitem{Gandhi:2004bj}
\bibinfo{author}{\bibfnamefont{R.}~\bibnamefont{Gandhi}},
  \bibinfo{author}{\bibfnamefont{P.}~\bibnamefont{Ghoshal}},
  \bibinfo{author}{\bibfnamefont{S.}~\bibnamefont{Goswami}},
  \bibinfo{author}{\bibfnamefont{P.}~\bibnamefont{Mehta}}, \bibnamefont{and}
  \bibinfo{author}{\bibfnamefont{S.~U.} \bibnamefont{Sankar}},
  \bibinfo{journal}{Phys.Rev.} \textbf{\bibinfo{volume}{D73}},
  \bibinfo{pages}{053001} (\bibinfo{year}{2006}), \eprint{hep-ph/0411252}.

\bibitem{Akhmedov:2004ny}
\bibinfo{author}{\bibfnamefont{E.~K.} \bibnamefont{Akhmedov}},
  \bibinfo{author}{\bibfnamefont{R.}~\bibnamefont{Johansson}},
  \bibinfo{author}{\bibfnamefont{M.}~\bibnamefont{Lindner}},
  \bibinfo{author}{\bibfnamefont{T.}~\bibnamefont{Ohlsson}}, \bibnamefont{and}
  \bibinfo{author}{\bibfnamefont{T.}~\bibnamefont{Schwetz}},
  \bibinfo{journal}{JHEP} \textbf{\bibinfo{volume}{0404}}, \bibinfo{pages}{078}
  (\bibinfo{year}{2004}), \eprint{hep-ph/0402175}.

\bibitem{Ribeiro:2007ud}
\bibinfo{author}{\bibfnamefont{N.}~\bibnamefont{Ribeiro}},
  \bibinfo{author}{\bibfnamefont{H.}~\bibnamefont{Minakata}},
  \bibinfo{author}{\bibfnamefont{H.}~\bibnamefont{Nunokawa}},
  \bibinfo{author}{\bibfnamefont{S.}~\bibnamefont{Uchinami}}, \bibnamefont{and}
  \bibinfo{author}{\bibfnamefont{R.}~\bibnamefont{Zukanovich-Funchal}},
  \bibinfo{journal}{JHEP} \textbf{\bibinfo{volume}{0712}}, \bibinfo{pages}{002}
  (\bibinfo{year}{2007}), \eprint{0709.1980}.

\bibitem{Kopp:2007ne}
\bibinfo{author}{\bibfnamefont{J.}~\bibnamefont{Kopp}},
  \bibinfo{author}{\bibfnamefont{M.}~\bibnamefont{Lindner}},
  \bibinfo{author}{\bibfnamefont{T.}~\bibnamefont{Ota}}, \bibnamefont{and}
  \bibinfo{author}{\bibfnamefont{J.}~\bibnamefont{Sato}},
  \bibinfo{journal}{Phys.Rev.} \textbf{\bibinfo{volume}{D77}},
  \bibinfo{pages}{013007} (\bibinfo{year}{2008}), \eprint{0708.0152}.

\bibitem{Blennow:2008eb}
\bibinfo{author}{\bibfnamefont{M.}~\bibnamefont{Blennow}} \bibnamefont{and}
  \bibinfo{author}{\bibfnamefont{T.}~\bibnamefont{Ohlsson}},
  \bibinfo{journal}{Phys. Rev.} \textbf{\bibinfo{volume}{D78}},
  \bibinfo{pages}{093002} (\bibinfo{year}{2008}), \eprint{0805.2301}.

\bibitem{Kikuchi:2008vq}
\bibinfo{author}{\bibfnamefont{T.}~\bibnamefont{Kikuchi}},
  \bibinfo{author}{\bibfnamefont{H.}~\bibnamefont{Minakata}}, \bibnamefont{and}
  \bibinfo{author}{\bibfnamefont{S.}~\bibnamefont{Uchinami}},
  \bibinfo{journal}{JHEP} \textbf{\bibinfo{volume}{0903}}, \bibinfo{pages}{114}
  (\bibinfo{year}{2009}), \eprint{0809.3312}.

\bibitem{Meloni:2009ia}
\bibinfo{author}{\bibfnamefont{D.}~\bibnamefont{Meloni}},
  \bibinfo{author}{\bibfnamefont{T.}~\bibnamefont{Ohlsson}}, \bibnamefont{and}
  \bibinfo{author}{\bibfnamefont{H.}~\bibnamefont{Zhang}},
  \bibinfo{journal}{JHEP} \textbf{\bibinfo{volume}{04}}, \bibinfo{pages}{033}
  (\bibinfo{year}{2009}), \eprint{0901.1784}.

\bibitem{Asano:2011nj}
\bibinfo{author}{\bibfnamefont{K.}~\bibnamefont{Asano}} \bibnamefont{and}
  \bibinfo{author}{\bibfnamefont{H.}~\bibnamefont{Minakata}},
  \bibinfo{journal}{JHEP} \textbf{\bibinfo{volume}{1106}}, \bibinfo{pages}{022}
  (\bibinfo{year}{2011}), \eprint{1103.4387}.

\bibitem{Coloma:2011rq}
\bibinfo{author}{\bibfnamefont{P.}~\bibnamefont{Coloma}},
  \bibinfo{author}{\bibfnamefont{A.}~\bibnamefont{Donini}},
  \bibinfo{author}{\bibfnamefont{J.}~\bibnamefont{Lopez-Pavon}},
  \bibnamefont{and} \bibinfo{author}{\bibfnamefont{H.}~\bibnamefont{Minakata}},
  \bibinfo{journal}{JHEP} \textbf{\bibinfo{volume}{1108}}, \bibinfo{pages}{036}
  (\bibinfo{year}{2011}), \eprint{1105.5936}.

\bibitem{raffeltbook}
\bibinfo{author}{\bibfnamefont{G.~G.} \bibnamefont{Raffelt}},
  \emph{\bibinfo{title}{Stars as Laboratories for Fundamental Physics: The
  Astrophysics of Neutrinos, Axions, and Other Weakly Interacting Particles}}
  (\bibinfo{publisher}{University of Chicago Press}, \bibinfo{year}{1996}).

\bibitem{Akhmedov:2006hb}
\bibinfo{author}{\bibfnamefont{E.~K.} \bibnamefont{Akhmedov}},
  \bibinfo{author}{\bibfnamefont{M.}~\bibnamefont{Maltoni}}, \bibnamefont{and}
  \bibinfo{author}{\bibfnamefont{A.~Y.} \bibnamefont{Smirnov}},
  \bibinfo{journal}{JHEP} \textbf{\bibinfo{volume}{0705}}, \bibinfo{pages}{077}
  (\bibinfo{year}{2007}), \eprint{hep-ph/0612285}.

\bibitem{Akhmedov:2008qt}
\bibinfo{author}{\bibfnamefont{E.~K.} \bibnamefont{Akhmedov}},
  \bibinfo{author}{\bibfnamefont{M.}~\bibnamefont{Maltoni}}, \bibnamefont{and}
  \bibinfo{author}{\bibfnamefont{A.~Y.} \bibnamefont{Smirnov}},
  \bibinfo{journal}{JHEP} \textbf{\bibinfo{volume}{0806}}, \bibinfo{pages}{072}
  (\bibinfo{year}{2008}), \eprint{0804.1466}.

\bibitem{Wolfenstein:1977ue}
\bibinfo{author}{\bibfnamefont{L.}~\bibnamefont{Wolfenstein}},
  \bibinfo{journal}{Phys. Rev.} \textbf{\bibinfo{volume}{D17}},
  \bibinfo{pages}{2369} (\bibinfo{year}{1978}).

\bibitem{Mikheev:1987qk}
\bibinfo{author}{\bibfnamefont{S.~P.} \bibnamefont{Mikheev}} \bibnamefont{and}
  \bibinfo{author}{\bibfnamefont{A.~Y.} \bibnamefont{Smirnov}},
  \bibinfo{journal}{Sov. Phys. Usp.} \textbf{\bibinfo{volume}{30}},
  \bibinfo{pages}{759} (\bibinfo{year}{1987}).

\bibitem{Gaisser:2002jj}
\bibinfo{author}{\bibfnamefont{T.~K.} \bibnamefont{Gaisser}} \bibnamefont{and}
  \bibinfo{author}{\bibfnamefont{M.}~\bibnamefont{Honda}},
  \bibinfo{journal}{Ann. Rev. Nucl. Part. Sci.} \textbf{\bibinfo{volume}{52}},
  \bibinfo{pages}{153} (\bibinfo{year}{2002}), \eprint{hep-ph/0203272}.

\bibitem{Chatterjee:2014vta}
\bibinfo{author}{\bibfnamefont{A.}~\bibnamefont{Chatterjee}},
  \bibinfo{author}{\bibfnamefont{K.}~\bibnamefont{Meghna}},
  \bibinfo{author}{\bibfnamefont{K.}~\bibnamefont{Rawat}},
  \bibinfo{author}{\bibfnamefont{T.}~\bibnamefont{Thakore}},
  \bibinfo{author}{\bibfnamefont{V.}~\bibnamefont{Bhatnagar}}, \emph{et~al.},
  \bibinfo{journal}{JINST} \textbf{\bibinfo{volume}{9}},
  \bibinfo{pages}{P07001} (\bibinfo{year}{2014}), \eprint{1405.7243}.

\bibitem{Chatterjee:2013qus}
\bibinfo{author}{\bibfnamefont{A.}~\bibnamefont{Chatterjee}},
  \bibinfo{author}{\bibfnamefont{P.}~\bibnamefont{Ghoshal}},
  \bibinfo{author}{\bibfnamefont{S.}~\bibnamefont{Goswami}}, \bibnamefont{and}
  \bibinfo{author}{\bibfnamefont{S.~K.} \bibnamefont{Raut}},
  \bibinfo{journal}{JHEP} \textbf{\bibinfo{volume}{1306}}, \bibinfo{pages}{010}
  (\bibinfo{year}{2013}), \eprint{1302.1370}.

\bibitem{Bueno:2007um}
\bibinfo{author}{\bibfnamefont{A.}~\bibnamefont{Bueno}},
  \bibinfo{author}{\bibfnamefont{Z.}~\bibnamefont{Dai}},
  \bibinfo{author}{\bibfnamefont{Y.}~\bibnamefont{Ge}},
  \bibinfo{author}{\bibfnamefont{M.}~\bibnamefont{Laffranchi}},
  \bibinfo{author}{\bibfnamefont{A.}~\bibnamefont{Melgarejo}}, \emph{et~al.},
  \bibinfo{journal}{JHEP} \textbf{\bibinfo{volume}{0704}}, \bibinfo{pages}{041}
  (\bibinfo{year}{2007}), \eprint{hep-ph/0701101}.

\bibitem{Barger:2012fx}
\bibinfo{author}{\bibfnamefont{V.}~\bibnamefont{Barger}},
  \bibinfo{author}{\bibfnamefont{R.}~\bibnamefont{Gandhi}},
  \bibinfo{author}{\bibfnamefont{P.}~\bibnamefont{Ghoshal}},
  \bibinfo{author}{\bibfnamefont{S.}~\bibnamefont{Goswami}},
  \bibinfo{author}{\bibfnamefont{D.}~\bibnamefont{Marfatia}}, \emph{et~al.},
  \bibinfo{journal}{Phys.Rev.Lett.} \textbf{\bibinfo{volume}{109}},
  \bibinfo{pages}{091801} (\bibinfo{year}{2012}), \eprint{1203.6012}.

\bibitem{Choubey:2015xha}
\bibinfo{author}{\bibfnamefont{S.}~\bibnamefont{Choubey}},
  \bibinfo{author}{\bibfnamefont{A.}~\bibnamefont{Ghosh}},
  \bibinfo{author}{\bibfnamefont{T.}~\bibnamefont{Ohlsson}}, \bibnamefont{and}
  \bibinfo{author}{\bibfnamefont{D.}~\bibnamefont{Tiwari}},
  \bibinfo{journal}{JHEP} \textbf{\bibinfo{volume}{12}}, \bibinfo{pages}{126}
  (\bibinfo{year}{2015}), \eprint{1507.02211}.

\end{thebibliography}

\end{document}